\documentclass[]{pasj02} 
\usepackage[switch,mathlines]{lineno} % add line number to manuscript
\usepackage{natbib} 
\usepackage{url}
\usepackage{amsmath}
\usepackage{zref-clever}
\usepackage{multirow}
\usepackage{comment}

\jyear{2024}
\Received{2026/02/18}%{yyyy/mm/dd}
\Accepted{2026/05/17}%{yyyy/mm/dd}
\newcommand{\satellite}{\text}
\newcommand{\filter}{\textit}
\newcommand{\footnoteurl}[1]{$\langle$\url{#1}$\rangle$}

\newcommand{\AU}{\text{AU}}
\newcommand{\kpc}{\text{kpc}}
\newcommand{\Kelvin}{\text{K}}
\newcommand{\Year}{\text{yr}}
\newcommand{\Days}{\text{d}}
\newcommand{\Minute}{\text{min}}
\newcommand{\Second}{\text{s}}
\newcommand{\EpS}{\text{erg}\>\Second^{-1}}
\newcommand{\EpSC}{\EpS\>\text{cm}^{-2}}
\newcommand{\kMpS}{\text{km}\>\Second^{-1}}
\newcommand{\keV}{\text{keV}}
\newcommand{\MJD}{\text{MJD}}
\newcommand{\lightsec}{\text{lt-s}}
\newcommand{\Deg}{\text{deg}}
\newcommand{\Crab}{\text{Crab}}
\newcommand{\Mag}{\text{mag}}

\newcommand{\Orb}{\mathrm{orb}}
\newcommand{\Star}{\mathrm{s}}
\newcommand{\Disk}{\mathrm{d}}
\newcommand{\Pulsar}{\mathrm{NS}}
\newcommand{\Opt}{\mathrm{opt}}
\newcommand{\Xray}{\mathrm{x}}

\newcommand{\diff}[1]{\D{#1}}
\newcommand{\Porb}{P_\Orb}
\newcommand{\phiorb}{\phi_\Orb}
\newcommand{\PhiOrb}{\Phi_\Orb}
\newcommand{\Pdot}{\dot{P}_\Orb/\Porb}
\newcommand{\fscf}{f_{0/4}}
\newcommand{\fleft}{f_{1/4}}
\newcommand{\fic}{f_{2/4}}
\newcommand{\fright}{f_{3/4}}
\newcommand{\fscl}{f_{4/4}}
\newcommand{\FbarX}{\bar{f}_\Xray}
\newcommand{\asymm}{\mathcal{A}}

\newcommand{\Mx}{M_\Pulsar}
\newcommand{\Ms}{M_\Star}

\newcommand{\logq}{\log{q}}
\newcommand{\iorb}{i_\Orb}
\newcommand{\Tstar}{T_\Star}
\newcommand{\Tdisk}{T_\Disk}
\newcommand{\aorb}{a}
\newcommand{\Lx}{L_\Xray}
\newcommand{\rdisk}{\hat{r}_\Disk}
\newcommand{\thetadisk}{\theta_\Disk}
\newcommand{\hdisk}{h_\Disk}
\newcommand{\hbdry}{h_\mathrm{b}}
\newcommand{\phidisk}{\phi_\Disk}
\newcommand{\phizero}{\phi_{\Disk,0}}
\newcommand{\phidot}{\dot{\phi}_\Disk}
\newcommand{\Fx}{F_\Xray}
\newcommand{\albedoS}{\alpha_\Star}
\newcommand{\albedoD}{\alpha_\Disk}

\newcommand{\fw}{f_\mathrm{w}}

\begin{document} 

\title{
    Optical Super-orbital Modulation of SMC X-1: Disk Precession and a Revised Pulsar Mass
}

%%% begin:list of authors
% Do NOT capitalize all letters in "textsc".
\author{
 Masafumi \textsc{Niwano}\altaffilmark{1}\altemailmark\orcid{0000-0003-3102-7452}, \email{masafumi.niwano@nao.ac.jp} 
 Nobuyuki \textsc{Kawai}\altaffilmark{2} and
 Michael \textsc{Fausnaugh} \altaffilmark{3} \orcid{0000-0002-9113-7162}
}
\altaffiltext{1}{
    National Astronomical Observatory of Japan, 2-21-1 Osawa, Mitaka, Tokyo 181-8588, Japan
}
\altaffiltext{2}{
    RIKEN, 2-1 Hirosawa, Wako, Saitama 351-0198, Japan
}
\altaffiltext{3}{
    Department of Physics and Astronomy, Texas Tech University, Lubbock TX, 79409-1051, USA
}

%%% end:list of authors

%% !!! Select 3 to 5 words from PASJ's key words !!! 
%% List of Key Words: https://academic.oup.com/pasj/pages/Pasj_Keywords 
%% "\KeyWords{ }" always has to be placed before ``\maketitle'' 

\maketitle

\begin{abstract}
The observational determination of the lower limit of neutron star masses is crucial for the physics of core-collapse supernovae.
In this light, SMC\,X-1 is an important object because of its estimated pulsar mass lying near or potentially below the theoretical lower limit.
SMC\,X-1 exhibits a double peaked optical orbital light curve due to the tidal distortion of the donor star, and analysis of this allows us to constrain the binary parameters.
In this study, we analyzed optical and X-ray light curves of SMC\,X-1 obtained by Transiting Exoplanet Survey Satellite and Monitor of All-sky X-ray Image.
We found the systematic variations in the optical orbital light curves synchronized with the X-ray super-orbital modulation, regarding the following two aspects: the minimum at inferior conjunction and the double-peak asymmetry.
To explain this behavior, we developed a modified ellipsoidal modulation model in which the precessing accretion disk changes the geometry of X-ray irradiation on the donor and that of optical irradiation on the disk.
As a result, this model succeeded in reproducing the observed optical and X-ray light curves.
Furthermore, we discovered that intense X-ray irradiation could cause the optical emission center to shift away from the gravitational center, potentially leading to an underestimation of the radial velocity of the donor by approximately 20\%.
Correcting for this effect yields an updated pulsar mass estimation of about $1.35\>\MO$.
\end{abstract}

%\pagewiselinenumbers 

\section{Introduction}
\label{sec:intro}
Observational determination of the allowed mass range of neutron stars (NSs) is crucial not only for astrophysics but also for more fundamental physics.
While the upper limit of NS masses is actively studied for the context of the equation of state for ultra-dense nuclear matter, the lower limit is also important, mainly for constraining the mechanism of core-collapse supernovae (CCSNe).
For example, \citet{strobel_1999} obtained the lower limit mass at $0.89\text{--}1.13\>\MO$ by analyzing stability of proto neutron stars during CCSNe.
Furthermore, recent CCSN simulations suggested a more stringent limit of $1.1\text{--}1.2\>\MO$ \citep{suwa_2018,nakamura_2025,muller_2025}.
This limit is primarily determined by the minimum Fe core mass for triggering a CCSN ($\approx$ Chandrasekhar mass) and by the energy loss due to neutrino radiation.

In this context, SMC\,X-1 is an important object, because of its pulsar with an estimated mass of $\lesssim1.1\>\MO$ \citep{vdm_2007,rawls_2011,coe_2013}.
SMC\,X-1 is a high-mass X-ray binary with an orbital period $\Porb=3.89\>\Days$, consisting of a pulsar with a period of $0.71\,\Second$ \citep{lucke_1976} and an early-type supergiant donor, Sk\,160 \citep{webster_1972}.
As indicated by its short $\Porb$, this is a very close binary system ($\aorb\sim0.1\>\AU$; where $\aorb$ is an orbital separation), and the donor is believed to be experiencing Roche-lobe overflow.
Furthermore, the X-ray luminosity of its pulsar reaches $\approx5\times10^{38}\>\EpS$ \citep{bonnet_1981,pike_2019}, exceeding the Eddington limit for NS masses, likely due to significant mass inflow via the Roche-lobe overflow.
The orbital period and projected semi-major axis of the pulsar orbit ($a_\mathrm{x}\sin{i}$) for SMC\,X-1 have been very precisely determined by pulse timing analyses \citep{inam_2010,raichur_2010}.
Therefore, the key to estimating its pulsar mass lies in the determination of the mass ratio and orbital inclination.
\citet{vdm_2007} derived the mass ratio of SMC\,X-1 by obtaining the radial velocity (RV) of the donor through spectroscopic observations, and determined the orbital inclination using the eclipse duration and the spherically approximated Roche-lobe.
Using these values, they estimated the pulsar mass as $1.06^{+0.11}_{-0.10}\>\MO$.
\citet{rawls_2011} employed the same basic methodology as \citet{vdm_2007} and referred to them for RV measurements, but additionally conducted a test of the spherical Roche-lobe approximation by fitting the orbital folded optical light curve with an ellipsoidal modulation model (described later).
Their estimated value for the pulsar mass is $1.04\pm0.09\>\MO$.
The mass estimation in \citet{coe_2013} was performed by applying the folded light curve analysis of \citet{rawls_2011} to data from another telescope, and obtained the value of $1.15\pm0.07\>\MO$.
In our study, we focus on two important features of SMC\,X-1: ellipsoidal modulation and super-orbital modulation.

\begin{figure}
    \includegraphics[width=\columnwidth]{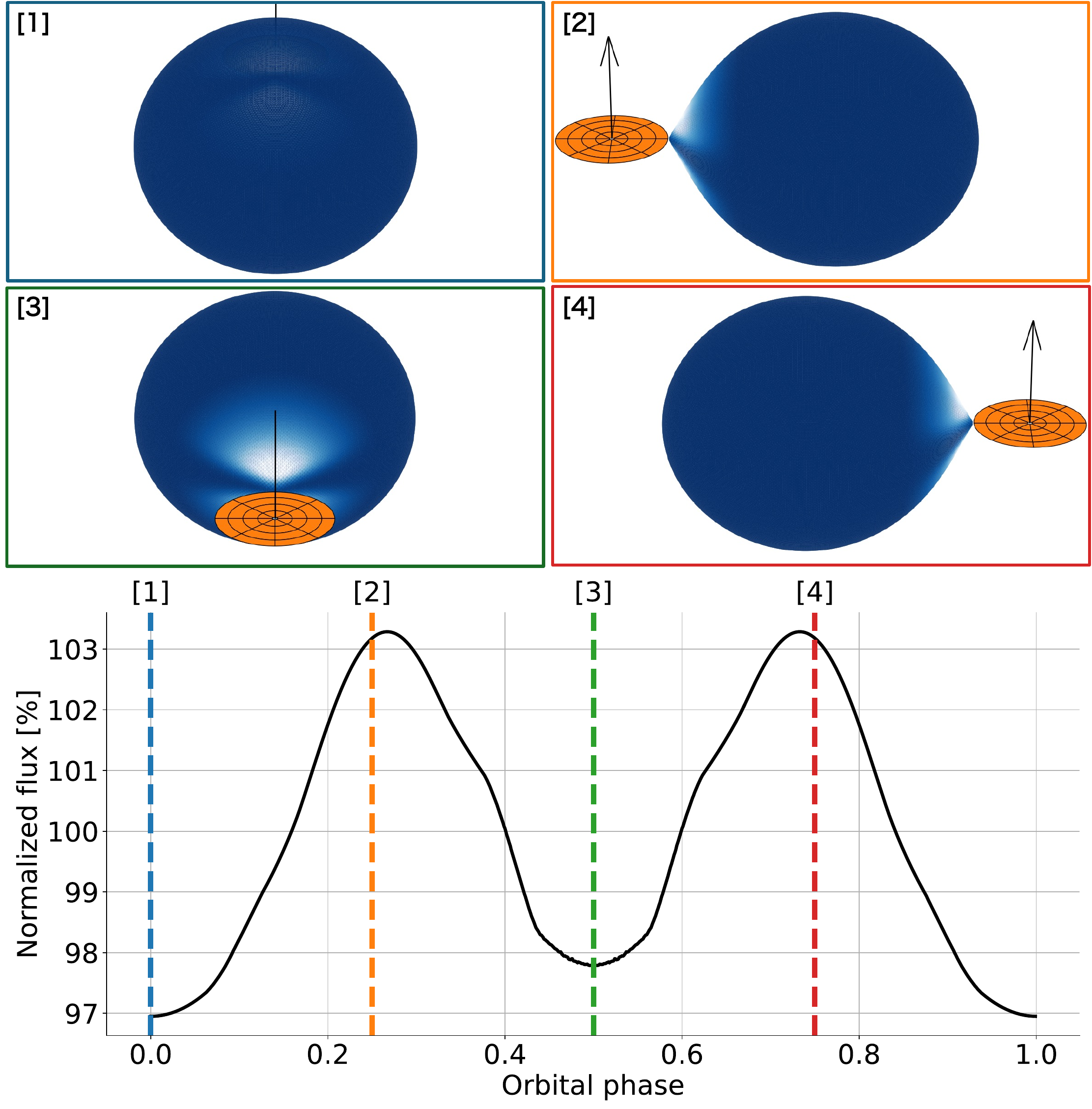}
    \caption{
        Schematic image of the ellipsoidal modulation of a high-mass X-ray binary.
        The top four panels are simulated images of the binary in the four distinctive phases, and the bottom panel is the light curve.
        The four vertical lines in the light curve correspond to the same numbered panel above.
        The teardrop-shaped object is the donor, the disk-shaped object is the accretion disk, and the arrow is the normal vector of the accretion disk.
        The color of the donor surface indicates the intensity of X-ray irradiation; the stronger the irradiation, the whiter the color.
        Note that the flux in the inferior conjunction is affected by a combination of various factors and may be smaller than that in the superior conjunction, unlike this figure.
        {
            Alt text: This figure is composed of simulated images of the binary in the four distinctive phases located in the upper half, and the orbital light curve in the lower half.
        }
    }
    \label{fig:ellipmod}
\end{figure}

An ellipsoidal modulation is a periodic and double-peaked optical flux variation, caused by the change in the apparent stellar size as it rotates while being deformed into a teardrop shape by the tidal forces of its companion (\zcref{fig:ellipmod}).
The shape of a deformed star is determined by the Roche potential, allowing the analysis of optical orbital light curves to constrain the mass ratio and orbital inclination.
An ellipsoidal modulation of SMC\,X-1 was confirmed by \citet{liller_1972}, and \citet{petro_1973b} made the first attempt to estimate the pulsar mass by analyzing its orbital light curve.
Since then, several similar studies have been carried out by previous researchers \citep{anvi_1975,rawls_2011,coe_2013}.

A super-orbital modulation is a (quasi-) periodic flux variation (generally in X-ray) with a longer period than the orbital period, observed in X-ray binaries.
SMC\,X-1 exhibits quasi-periodic super-orbital modulation with a $40\text{--}60\>\Days$ period \citep{gruber_1984,wojdowski_1998}.
The most widely accepted interpretation of these phenomena is periodic occultation of the central X-ray source by a precessing accretion disk.
This physical picture was originally developed to explain the super-orbital modulation of Her\,X-1 \citep{katz_1973,petterson_1975,schwarzenberg_1992,wijers_1999}, and has since been applied to other systems, including SMC\,X-1.
Super-orbital modulations are not limited to the X-ray band, and ``optical'' super-orbital modulations---optical variations synchronized with the X-ray super-orbital cycle---have been reported in several X-ray binaries, such as Her\,X-1, LMC\,X-4, and SS\,433 \citep{petro_1973a,chevalier_1981,henson_1983}.
SMC\,X-1 also belongs to this class, with reports by \citet{coe_2013} and \citet{hu_2019}.
However, these two studies were limited to only confirming variations in the folded and averaged optical light curve by the super-orbital period.
This is likely due to the use of ground-based optical data, which suffer from restricted photometric precision and sparse temporal sampling.
In the cases of Her\,X-1 \citep{gerand_1976} and LMC\,X-4 \citep{heemsherk_1989}, optical super-orbital modulation has been understood as a consequence of disk precession changing the X-ray irradiation on the donor, the apparent disk size, and the geometry of the donor occultation.
In these previous studies, the limitations of ground-based observations were mitigated by averaging optical orbital light curves observed at the same super-orbital phase.

Reports of the optical super-orbital modulation in SMC\,X-1 cast doubt on its previous pulsar mass estimates based on the analysis of folded orbital light curves.
However, with ground-based observations, it is difficult to detect its optical super-orbital modulation without super-orbital folding and averaging, making it challenging to investigate the details of the optical super-orbital modulation and modify the analysis of the ellipsoidal modulation.
Transiting Exoplanet Survey Satellite (\satellite{TESS}) can be a viable solution to this problem, because it possesses the sufficient observational cadence and photometric precision to investigate the optical light curves of SMC\,X-1 for each orbital cycle.
Furthermore, combining this with long-term X-ray monitoring data from Monitor of All-sky X-ray Image (MAXI) enables the tracking of the X-ray super-orbital modulation.

The objective of this study is to determine the pulsar mass of SMC\,X-1 observationally.
For addressing this, we analyzed optical and X-ray light curves of \satellite{TESS} and MAXI, and built a modified ellipsoidal modulation code in which the precessing accretion disk changes both the pattern of X-ray irradiation on the donor surface and the optical reprocessing on the disk.
\zcref[S]{sec:obs} describes observations by \satellite{TESS} and MAXI.
Next, we show the observed optical and X-ray orbital light curves, and the detail of optical super-orbital modulations we confirmed, in \zcref{sec:optical_so_modulation}.
In \zcref{sec:modified_ellipmod_model}, we develop a modified ellipsoidal modulation code to reproduce the optical super-orbital modulation, and perform Markov chain Monte Carlo sampling for fitting the model to the light curves and estimation of binary parameters.
Finally, \zcref{sec:summary} summarizes this study.
Unless otherwise specified, this paper describes $1\sigma$ range as numerical uncertainty.

\section{Observations}
\label{sec:obs}
\subsection{\satellite{TESS}}
\label{sec:tess}
\satellite{TESS} \citep{ricker_2015} is a space telescope equipped with four wide-field optical cameras, operated by the National Aeronautics and Space Administration (NASA) and the Massachusetts Institute of Technology.
The field where \satellite{TESS} observes in one observational epoch is called a `sector'.
We used full frame images (FFIs) with intervals of $200\>\Second$ to $30\>\Minute$ in six sectors, which contain the SMC\,X-1 field.
\zcref[S]{tab:sectors} describes these six sectors.

\begin{table}
  \tbl{
    Sectors of \satellite{TESS} data we used in this study 
   }{
      \begin{tabular}{cccc}
        \hline
        Sector & Start & End & FFI intervals\\
        & [\MJD] & [\MJD] &\\
        \hline
        1 & 58324 & 58352 & 30\>\Minute\\
        2 & 58352 & 58381 & 30\>\Minute\\
        13 & 58653 & 58682 & 30\>\Minute\\
        27 & 59034 & 59060 & 10\>\Minute\\
        28 & 59060 & 59087 & 10\>\Minute\\
        68 & 60154 & 60181 & 200\>\Second\\
        \hline
      \end{tabular}
  }
  \label{tab:sectors}
  \begin{tabnote}
  \end{tabnote}
\end{table}

For the photometry of FFIs, we employed the photometry pipeline of \satellite{TESS} Transients project \citep{fausnaugh_2021,fausnaugh_2023}, which is based on image differencing and point-spread-function fitting.
This approach is expected to be more accurate than a simple aperture photometry for fainter sources ($\gtrsim12\>\Mag$ in \satellite{TESS} band).
We converted the photoelectron count rate recorded in FFIs to an energy flux density according to the documentation of \satellite{TESS} Transients\footnote{\footnoteurl{https://tess.mit.edu/public/tesstransients}}.
Because the photometry is applied to difference images, the measured flux represents deviations from the average flux level rather than the absolute flux.
We added a constant baseline flux to these difference light curves.
This procedure implicitly assumes that the baseline flux of SMC\,X-1 is constant across all sectors; however, the baseline flux may vary by up to $\sim3\%$ depending on the super-orbital phase \citep{coe_2013}.
Therefore, the optical super-orbital modulation of \citet{coe_2013} may be suppressed in our light curves.
The baseline flux is calculated from the \satellite{Gaia}-\filter{RP} band magnitude ($13.319\pm0.07$; \citealp{gaia_2023}) based on similarity between the \satellite{TESS} and \satellite{Gaia}-\filter{RP} filters.

\subsection{MAXI}
\label{sec:maxi}
MAXI \citep{matsuoka_2009} is an X-ray all-sky camera installed onboard International Space Station.
We obtained $90\>\Minute$-binned Gas Slit Camera (GSC) $2\text{--}20\>\keV$ light curves covering a modified Julian date (MJD) range of $58250\text{--}60250$, using MAXI on-demand web interface \citep{nakahira_2012}.
As a rough estimate of the energy flux, we calculated the flux in the unit of $\Crab$ by comparing to the GSC $2\text{--}20\>\keV$ count rate of the Crab nebula.

\section{Light curves and their characteristics}
\label{sec:optical_so_modulation}
We calculated the orbital phase referring to the orbital ephemeris of \citet{hu_2019}, which is derived by the analysis of data from MAXI, \satellite{Swift}, \satellite{RXTE}, \satellite{Chandra}, and \satellite{XMM-Newton}: $T_{\pi/2,0}=52846.6913(2)\>\MJD$, $P_{\Orb,0}=3.8919297(2)\>\Days$, and $\Pdot=-3.380(6)\times10^{-6}\>\Year^{-1}$, where $T_{\pi/2,0}$ is the time of the reference superior conjunction\footnote{The subscript $\pi/2$ means the angle of rotation from the intersection line of orbital and celestial plane.} and $P_{\Orb,0}$ is the orbital period at $T_{\pi/2,0}$.
The cumulative orbital phase $\PhiOrb$ is first obtained by integrating the reciprocal of $\Porb$ over time starting from $T_{\pi/2,0}$.
Its integer part corresponds to the number of orbital cycles counting from $T_{\pi/2,0}$, while the fractional part represents the orbital phase $\phiorb$ within that cycle.
In this paper, we use the integer part of $\PhiOrb$ as an identifier of each orbital cycle.

\subsection{Optical and X-ray light curves}
\label{sec:lightcurves}
\zcref[S]{fig:lc_sector01-13,fig:lc_sector27-68} are the optical and X-ray light curves of SMC\,X-1 in sectors 1--68.
Ellipsoidal modulation in optical orbital light curves, periodic X-ray dip due to pulsar eclipse, and X-ray super-orbital modulation can be confirmed.
The cumulative orbital phase $\PhiOrb$ derived from \citet{hu_2019} was in agreement with both optical and X-ray light curves for all the six sectors.

\begin{figure}
    \includegraphics[width=\columnwidth]{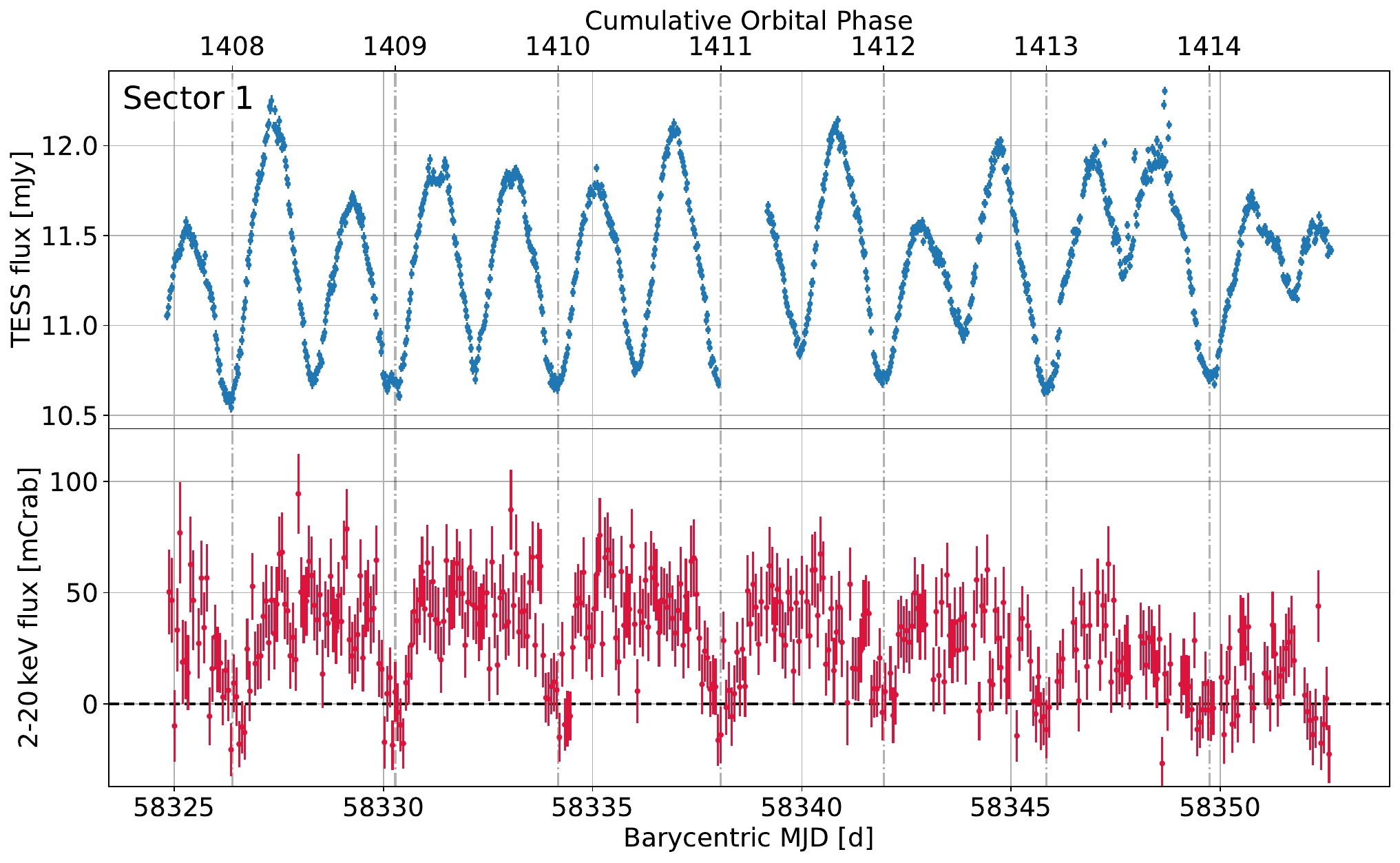}
    \includegraphics[width=\columnwidth]{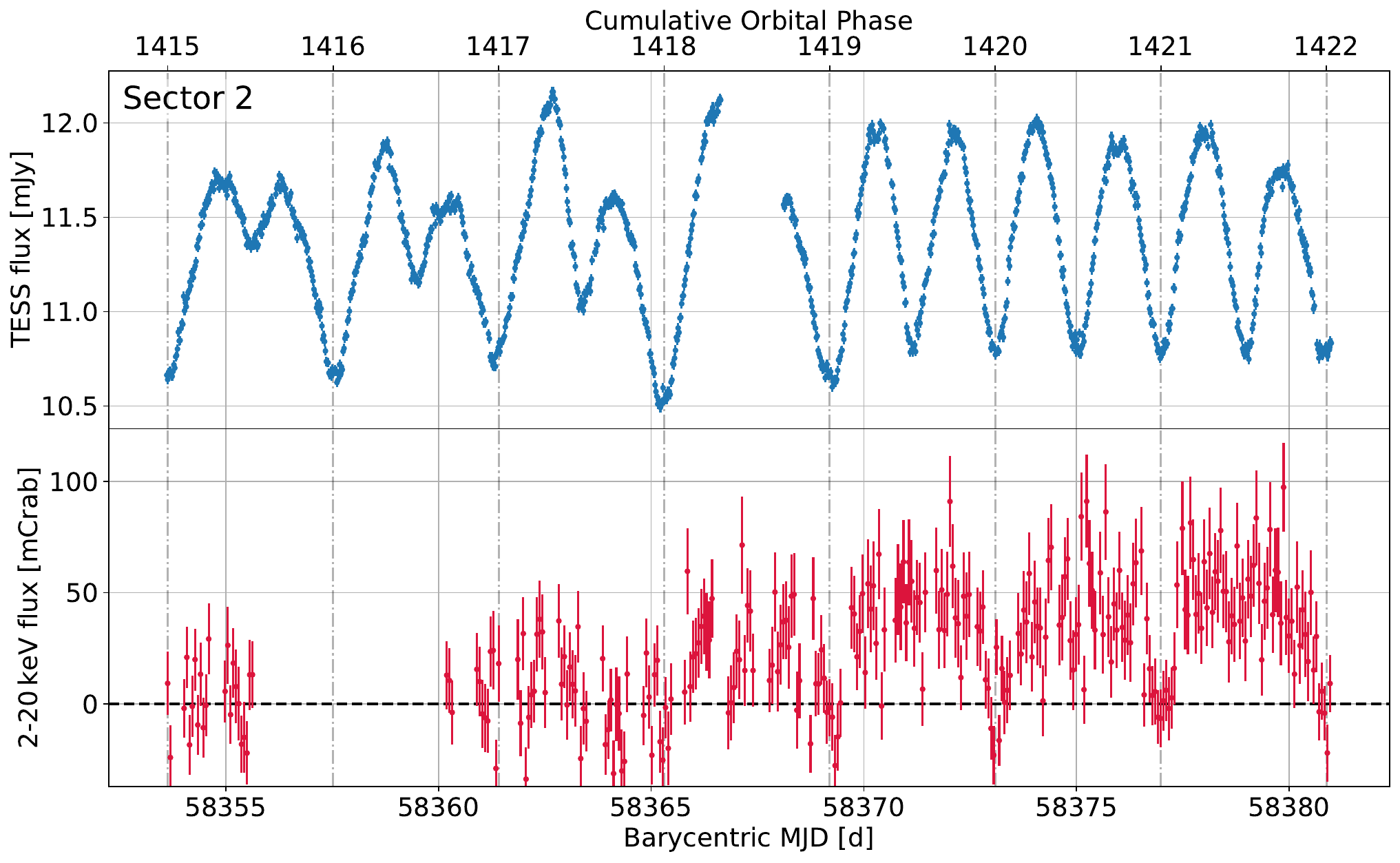}
    \includegraphics[width=\columnwidth]{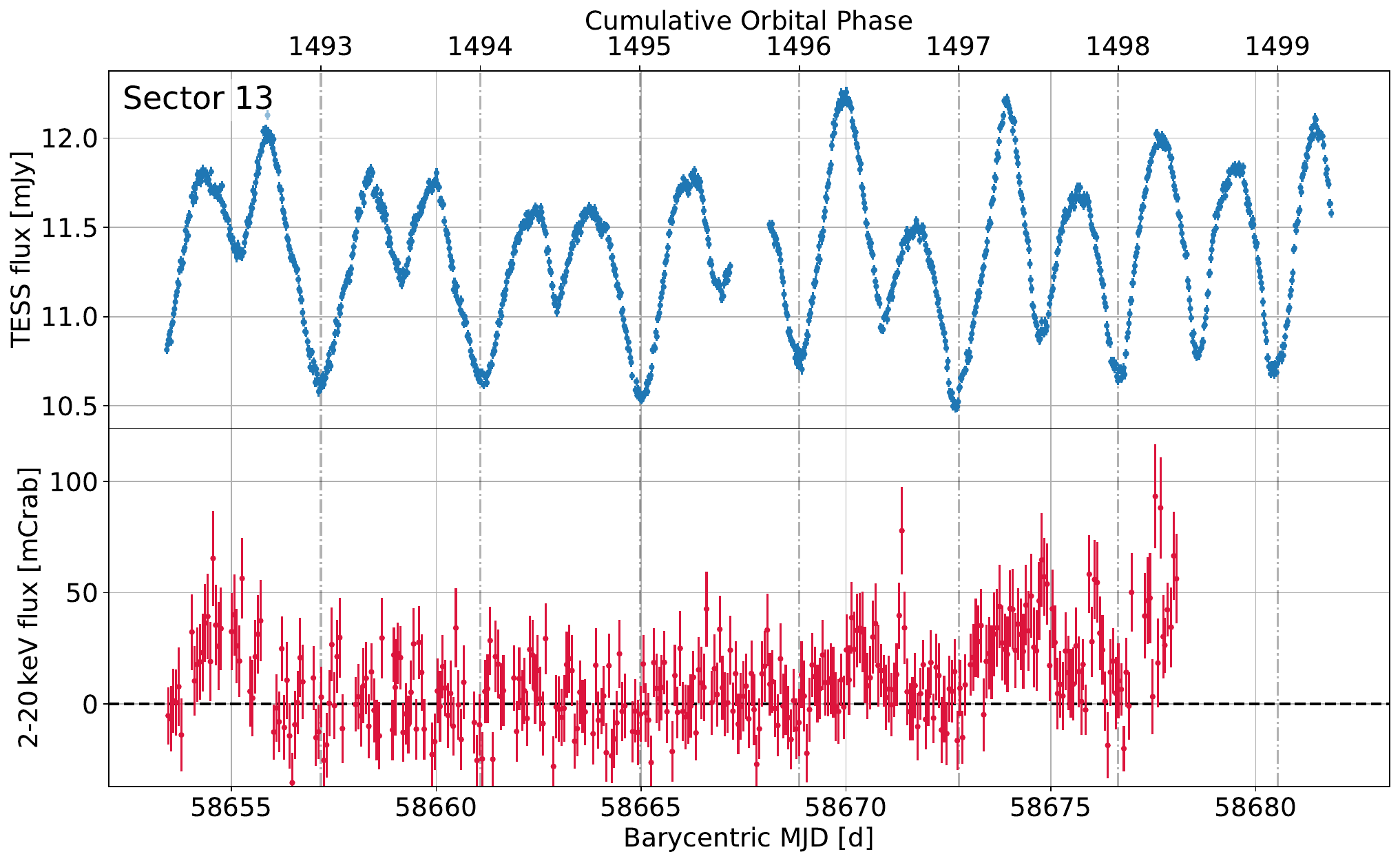}
    \caption{
        Optical and X-ray light curves in sectors 1, 2, and 13
        {
            Alt text: This figure consists of six panels arranged vertically, with each pair of adjacent panels forming a set.
            The upper panel of the pair shows the light curve for visible light, while the lower panel shows the light curve for X-rays.
            The three pairs correspond to the three sectors.
            For each pair, the upper and lower panels show the optical and X-ray light curves, respectively.
        }
    }
    \label{fig:lc_sector01-13}
\end{figure}
\begin{figure}
    \includegraphics[width=\columnwidth]{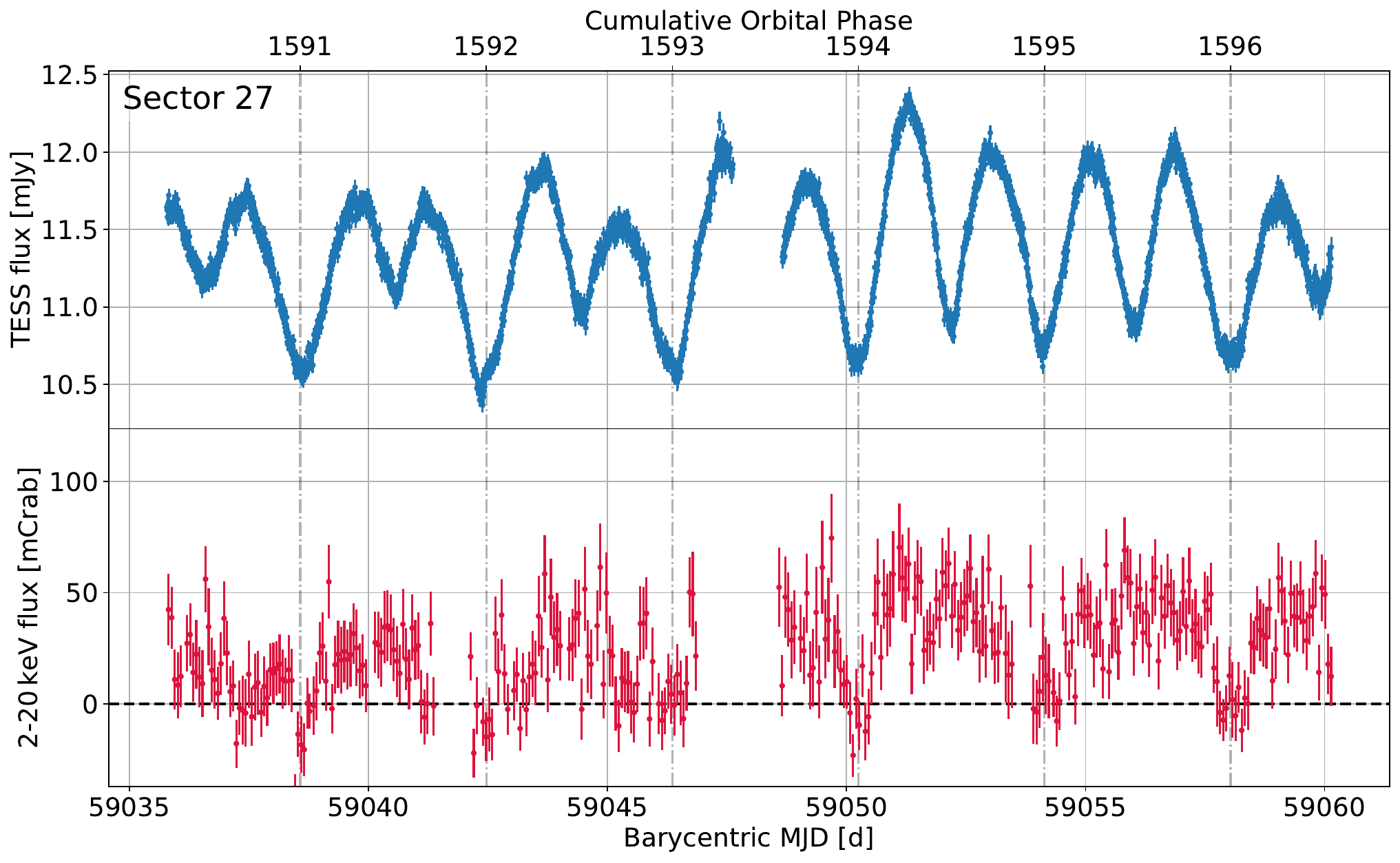}
    \includegraphics[width=\columnwidth]{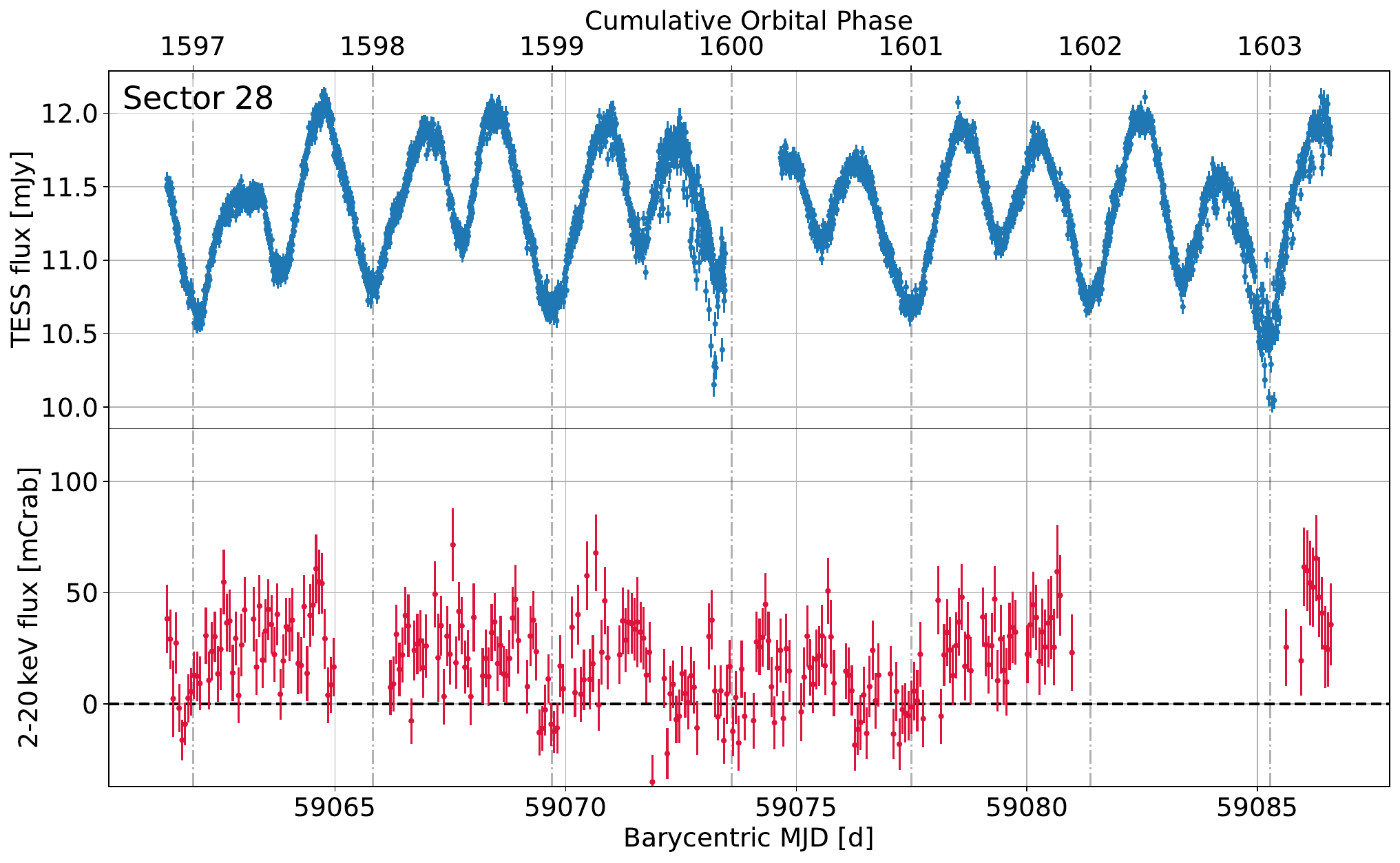}
    \includegraphics[width=\columnwidth]{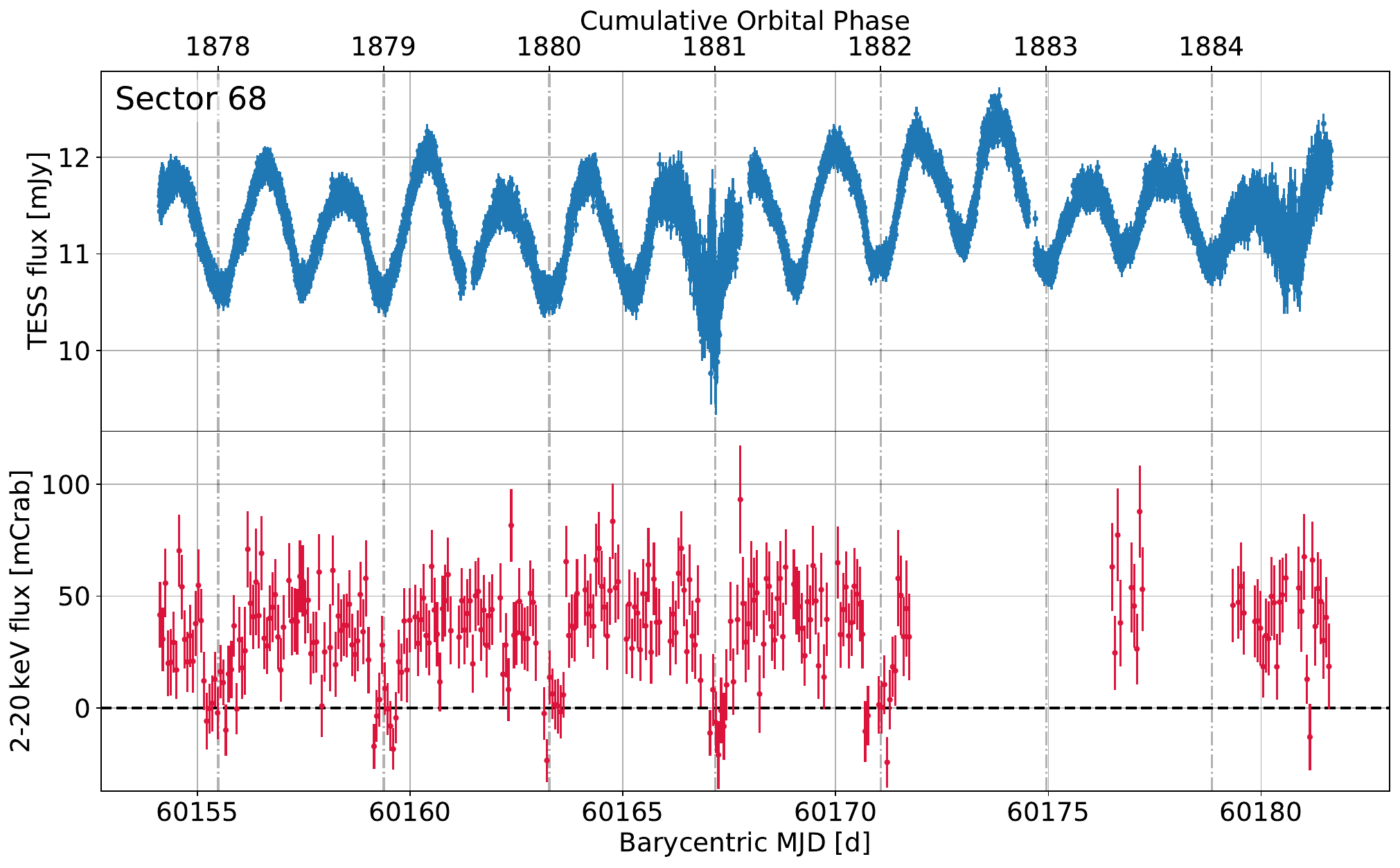}
    \caption{
        Optical and X-ray light curves in sectors 27, 28, and 68
        {
            Alt text: Same configuration as the figure of light curves for sectors 1, 2, and 13.
        }
    }
    \label{fig:lc_sector27-68}
\end{figure}

It is worth noting that the shape of the optical light curve systematically varied with each orbit in the following two aspects: (1) the depth of the minimum at the inferior conjunction ($\phiorb\approx1/2$), and (2) the asymmetry in the double peak heights.
This cannot be explained by a simple and static ellipsoidal modulation model.
In addition, these optical variations appear to be linked to the X-ray super-orbital modulation.

\subsection{Optical and X-ray correlations}
\label{sec:ox_correlation}
\begin{figure}
    \includegraphics[width=\columnwidth]{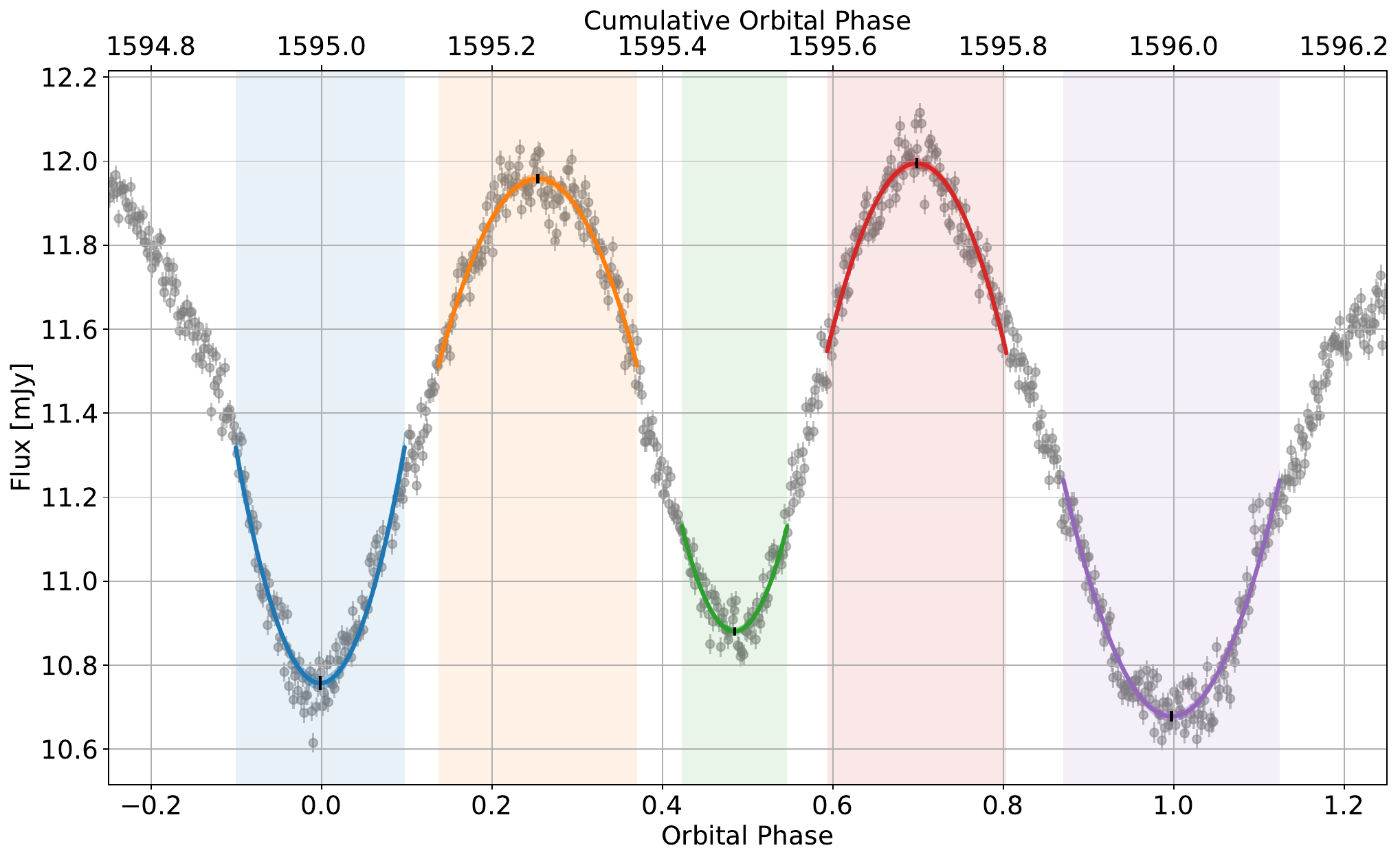}
    \caption{
        Parabola fitting of orbital light curves at minimum and peak phases.
        The flux at each parabola peak is labeled $\fscf$, $\fleft$, $\fic$, $\fright$, and $\fscl$, from left to right, respectively.
        {
            Alt text: Using the light curve of cycle 1595 as an example, it shows the observed TESS light curve and the five parabolic curves obtained from fitting at five phases.
        }
    }
    \label{fig:parabola_fit}
\end{figure}

\begin{figure}
    \includegraphics[width=\columnwidth]{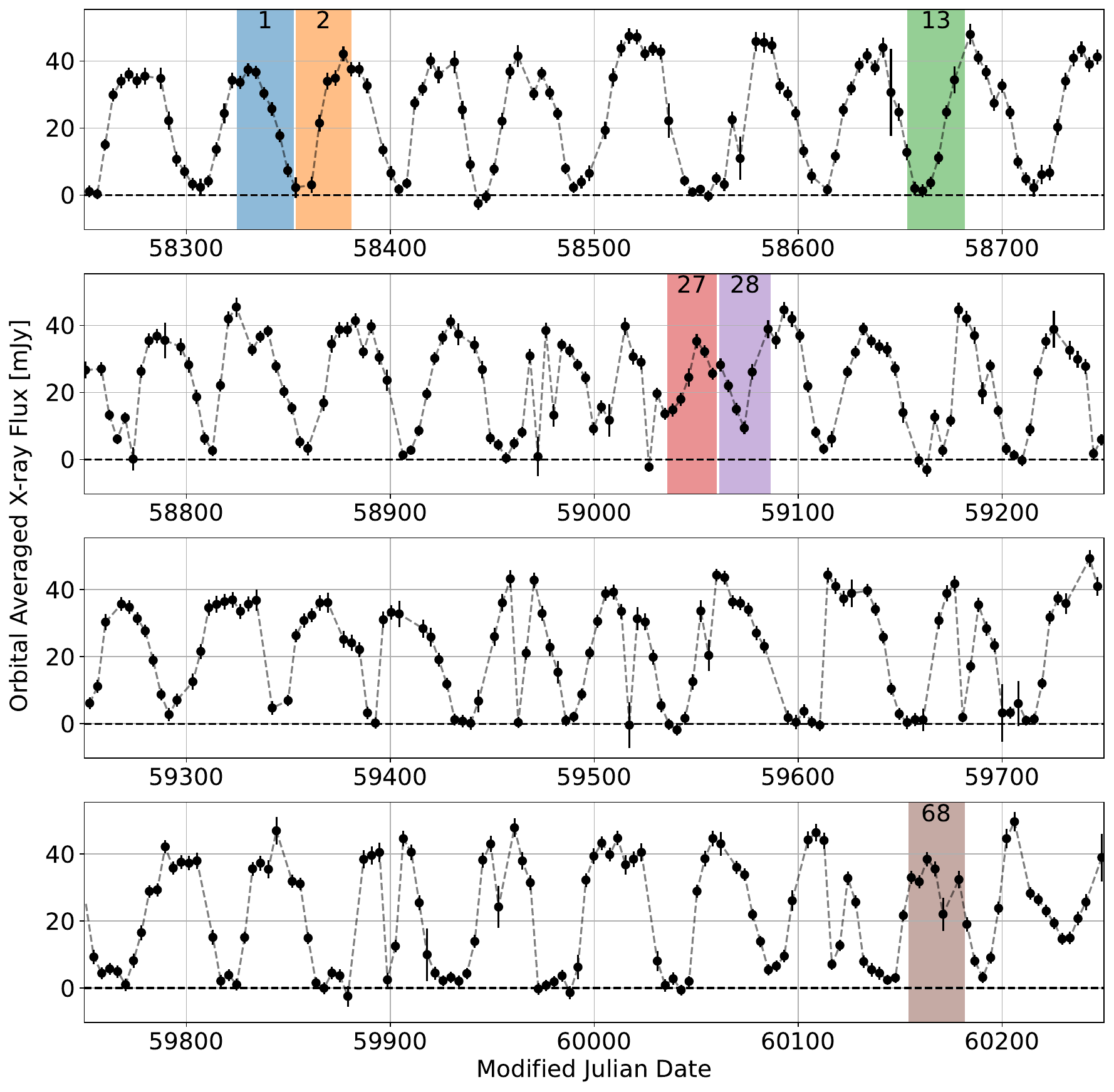}
    \caption{
        Long-term orbital-averaged X-ray light curves.
        Shaded areas indicate the sectors of \satellite{TESS} observations.
        {
            Alt text: The orbital-averaged X-ray light curve for the our MAXI dataset is shown in four panels arranged vertically for the time span MJD 58250--60250 divided into four segments.
        }
    }
    \label{fig:xray_superorbital}
\end{figure}

We use fluxes at the minimum and peak phases of the optical orbital light curve for evaluating the optical super-orbital modulation.
The orbital light curves at the peak and minimum phases were fitted with parabolas (\zcref{fig:parabola_fit}).
The procedure for this parabola fitting is described in \zcref{sec:parabola_fit}.
In this paper, the fluxes at the left superior conjunction ($\phiorb=0$), left peak ($\phiorb\approx1/4$), inferior conjunction ($\phiorb\approx1/2$), right peak ($\phiorb\approx3/4$), and right superior conjunction ($\phiorb=1$) are referred to as $\fscf$, $\fleft$, $\fic$, $\fright$, and $\fscl$, respectively.
Note that $\fscl$ is the same as $\fscf$ of the next orbital cycle.
We used a difference over sum ratio of double peak heights $\asymm$ as a measure of the double peak asymmetry:
\begin{equation}
    \asymm\equiv\frac{\fleft-\fright}{\fleft+\fright}.
\end{equation}
Furthermore, to evaluate the X-ray super-orbital modulation, the $2\text{--}20\>\keV$ flux was averaged over each orbital cycle, and we refer it as $\FbarX$.
The derivative $\diff{\FbarX}/\diff{t}$ was calculated by dividing the difference in $\FbarX$ between the front and rear points by the time interval between them.
We show long-term variations of $\FbarX$ in \zcref{fig:xray_superorbital}.
Quasi-periodic X-ray variations are clearly evident around sectors 1, 2, and 13, while $\FbarX$ exhibits complex variations around sectors 27, 28, and 68.
Such irregular super-orbital variations may be related to the ``excursion'' events \citep{wojdowski_1998,dage_2019}.

\zcref[S]{fig:opt-X,fig:peak-assymetry} show correlations between $\FbarX$ and minimum fluxes, $\FbarX$ and the peak flux, and $\diff{\FbarX}/\diff{t}$ and $\asymm$, respectively.
The uncertainties in the correlation coefficients shown in these figures were estimated using a combination of Monte Carlo error propagation and nonparametric bootstrap resampling.
Slightly stronger correlations are shown between $\FbarX$ and $\fic$, and $\diff{\FbarX}/\diff{t}$ and $\asymm$ with correlation coefficients of $-0.80\pm0.08$ and $0.63\pm0.10$, respectively.
In other words, (1) the minimum is shallow when X-rays are dark (off state) and deep when X-rays are bright (on state), and (2) the left (right) peak is larger during X-ray rising (decaying).
These correlations are also evident within each sector in the light curves (\zcref{fig:lc_sector01-13,fig:lc_sector27-68}).
Notably, all six sectors follow the same diagonal trend, which is remarkable given that the X-ray super-orbital modulation shows irregular behavior in sectors 27, 28, and 68.
There also is a modest positive correlation between $\FbarX$ and $(\fleft+\fright)/2$, average double peak flux.
On the other hand, the average minimum flux in the superior conjunction, $(\fscf+\fscl)/2$, did not correlate with the X-ray super-orbital modulation and was generally constant.

\begin{figure}
    \includegraphics[width=\columnwidth]{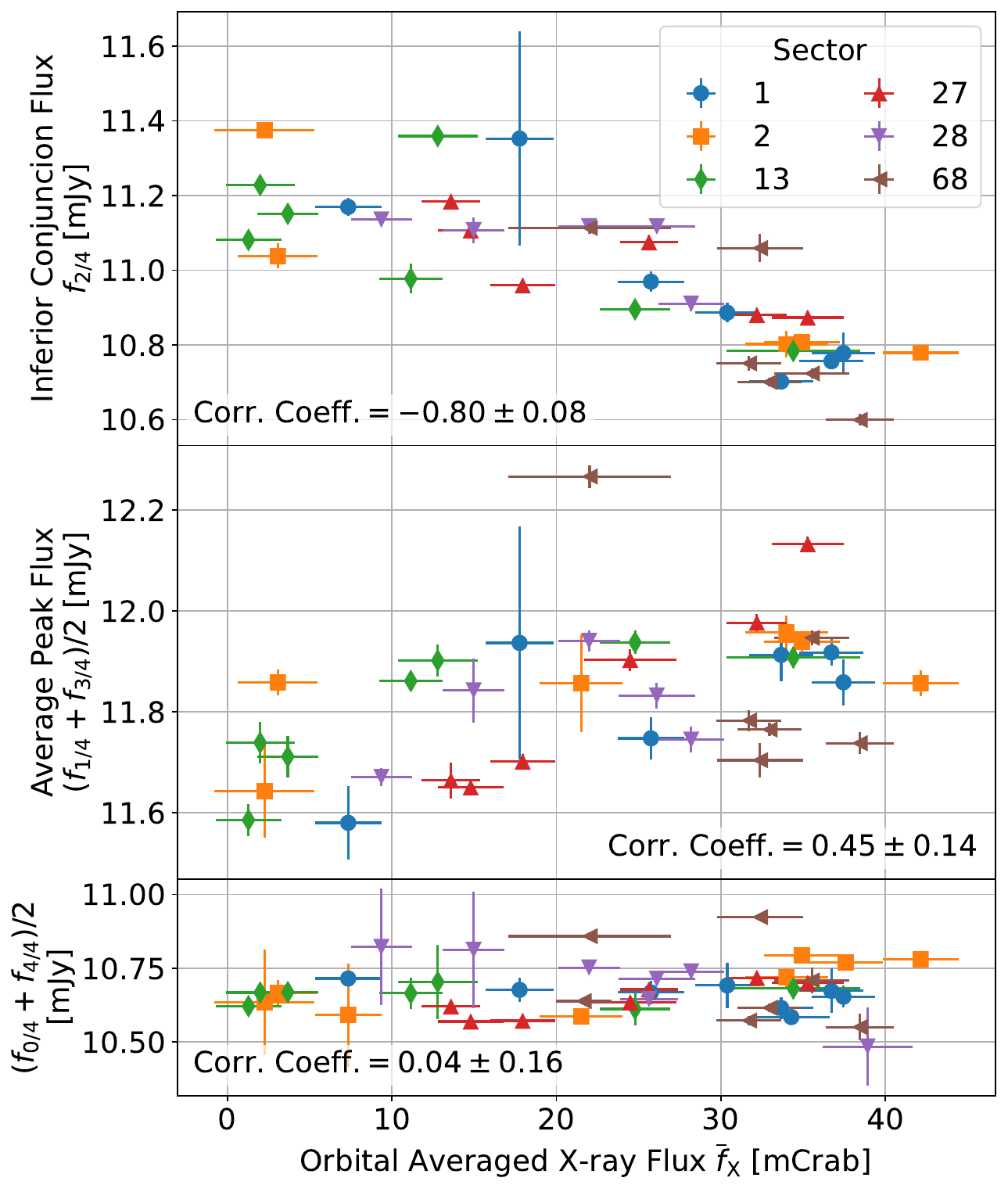}
    \caption{
        Correlations between optical and X-ray fluxes.
        The three panels show scatter plots with the Y-axis representing, from top to bottom, the inferior conjunction flux, the average of double peak heights, and the superior conjunction flux, respectively, while the X-axis represents the orbital-averaged X-ray flux for all three.
        {
            Alt text: A three-panel figure showing the correlations between optical fluxes at several orbital phases and the orbital-averaged X-ray flux.
        }
    }
    \label{fig:opt-X}
\end{figure}
\begin{figure}
    \includegraphics[width=\columnwidth]{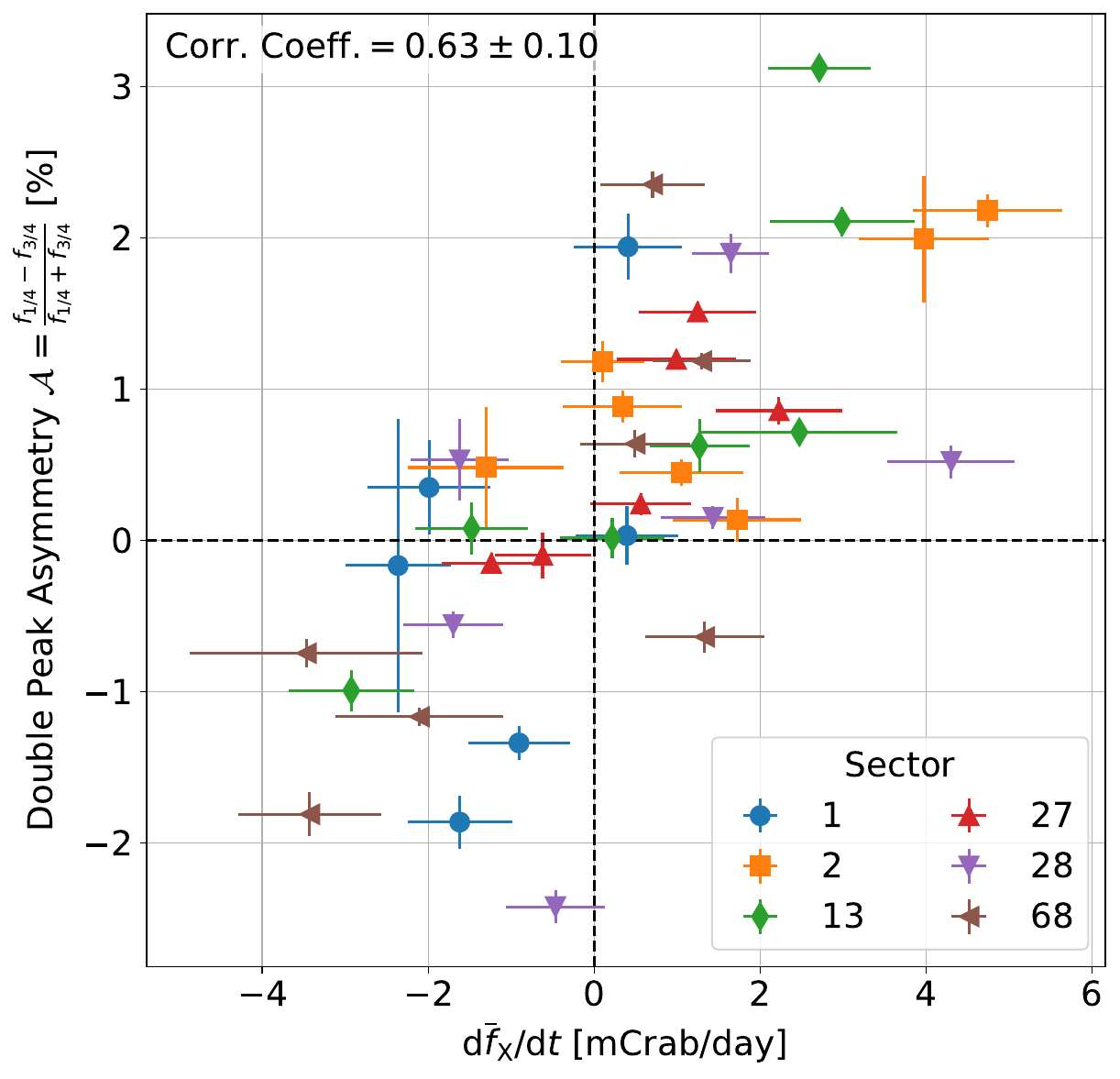}
    \caption{
        Scatter plot of X-ray super-orbital transition rate and the asymmetry of double peak heights.
        {
            Alt text: Scatter plot of X-ray super-orbital transition rate versus the asymmetry of double peak heights.
        }
    }
    \label{fig:peak-assymetry}
\end{figure}

\section{Modified ellipsoidal modulation model}
\label{sec:modified_ellipmod_model}
The optical super-orbital modulation exhibits notable correlations with the X-ray super-orbital modulation, except around the superior conjunction phase (i.e., during the pulsar eclipse).
This fact strongly suggests that the origin of the optical super-orbital modulation lies in regions that are also occulted during the eclipse, such as the accretion disk or the X-ray irradiated surface of the donor.
We propose a scenario in which the precessing accretion disk causes optical super-orbital modulation, through variations in the X-ray irradiation on the donor surface and in the reprocessed optical emission on the disk.
This follows the basic ideas of \citet{gerand_1976} on Her\,X-1 and \citet{heemsherk_1989} on LMC\,X-4, but differs in that our model takes into account the reprocessing of optical and ultraviolet (OUV) flux from the donor on the disk surface.
An additional notable difference is that our work reproduces not only the optical light curve but also the super-orbital X-ray light curve in the subsequent light curve fitting.

To describe the binary geometry, we define a Cartesian coordinate system centered on the donor: the $X$-axis points toward the pulsar, the $Z$-axis is parallel to the orbital axis and directed toward the observer, and the $Y$-axis lies in the orbital plane, $90^\circ$ counterclockwise from the $X$-axis.
The disk tilt direction is defined as the direction of the upward normal vector of the disk projected onto the $X$-$Y$ plane.
In the Roche-lobe derivation, we additionally use a spherical coordinate system $(r, \theta, \phi)$ defined in the standard way (i.e., $r$ is a radial distance from the origin, $\theta$ is a colatitude from the $Z$-axis, and $\phi$ is an azimuthal angle from the $X$-axis).
Furthermore, we assumed the counterclockwise orbiting as seen from the observer.

\subsection{Model implementation}
\subsubsection{Basic concept of the model}
\label{sec:model_concept}
\begin{figure}
    \includegraphics[width=\columnwidth]{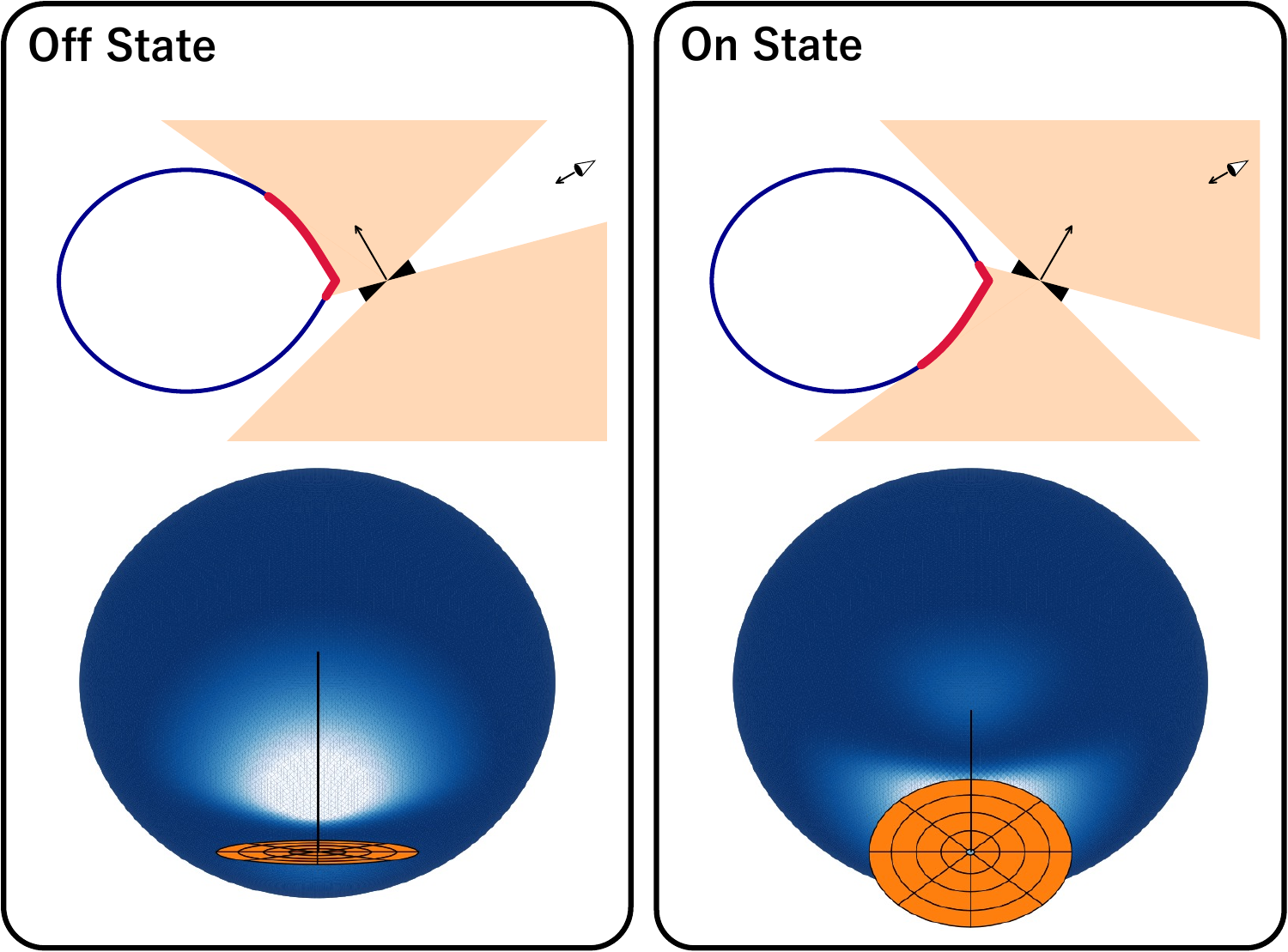}
    \caption{
        Schematic images of the binary with the misaligned accretion disk at the inferior conjunction phase in X-ray off and on state.
        The upper figure is a cross-sectional view in the $X$-$Z$ plane, where the teardrop-shaped object is the outline of the donor, the pair of wedges facing each other is the accretion disk, and the arrow extending from it is the normal vector of the disk.
        The irradiated area on the donor surface is showed as a thick red line.
        The object in the upper right corner indicates the direction in which the observer is located.
        The area where X-rays from the pulsar are not occluded by either the disk or the donor is shaded.
        The lower panel is a simulated image viewed from the observer.
        {
            Alt text: This compares X-ray irradiation on the donor when the disk is near edge-on and when it is tilted toward the observer.
        }
    }
    \label{fig:disk_tilt1}
\end{figure}

\zcref[S]{fig:disk_tilt1} shows the geometry of the binary with the misaligned accretion disk at inferior conjunction, in the X-ray off and on state.
In the off state (the left panel on \zcref{fig:disk_tilt1}), fewer X-rays reach the observer because of the (nearly) edge-on disk, while the pulsar irradiates the upper donor hemisphere, enhancing the optical emission.
In contrast, in the on state (the right panel), X-rays illuminate the lower hemisphere at inferior conjunction, resulting in smaller optical flux due to the reduced apparent irradiated area and the occultation by the disk.

\begin{figure}
    \includegraphics[width=\columnwidth]{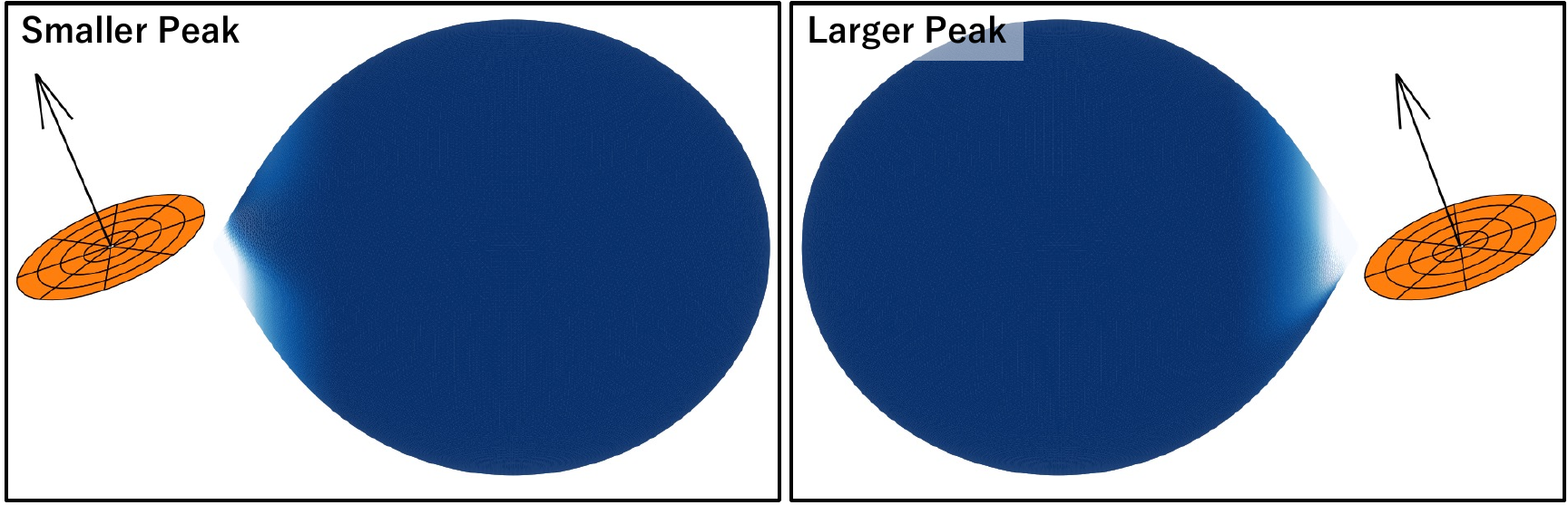}
    \caption{
        Simulated images of the binary with the misaligned disk viewed from the observer, at each of the double peak of the optical orbital light curve during the super-orbital transition phase.
        This figure shows a case where the disk is tilted toward the left side of the diagram; the first peak (on the left in the light curve) is small, while the second peak (on the right) is large.
        {
            Alt text: This compares X-ray irradiation on the donor when the disk is tilted toward the donor or the opposite, when the donor and disk are side-by-side as seen by an observer.
        }
    }
    \label{fig:disk_tilt2}
\end{figure}

\begin{figure*}
    \includegraphics[width=\linewidth]{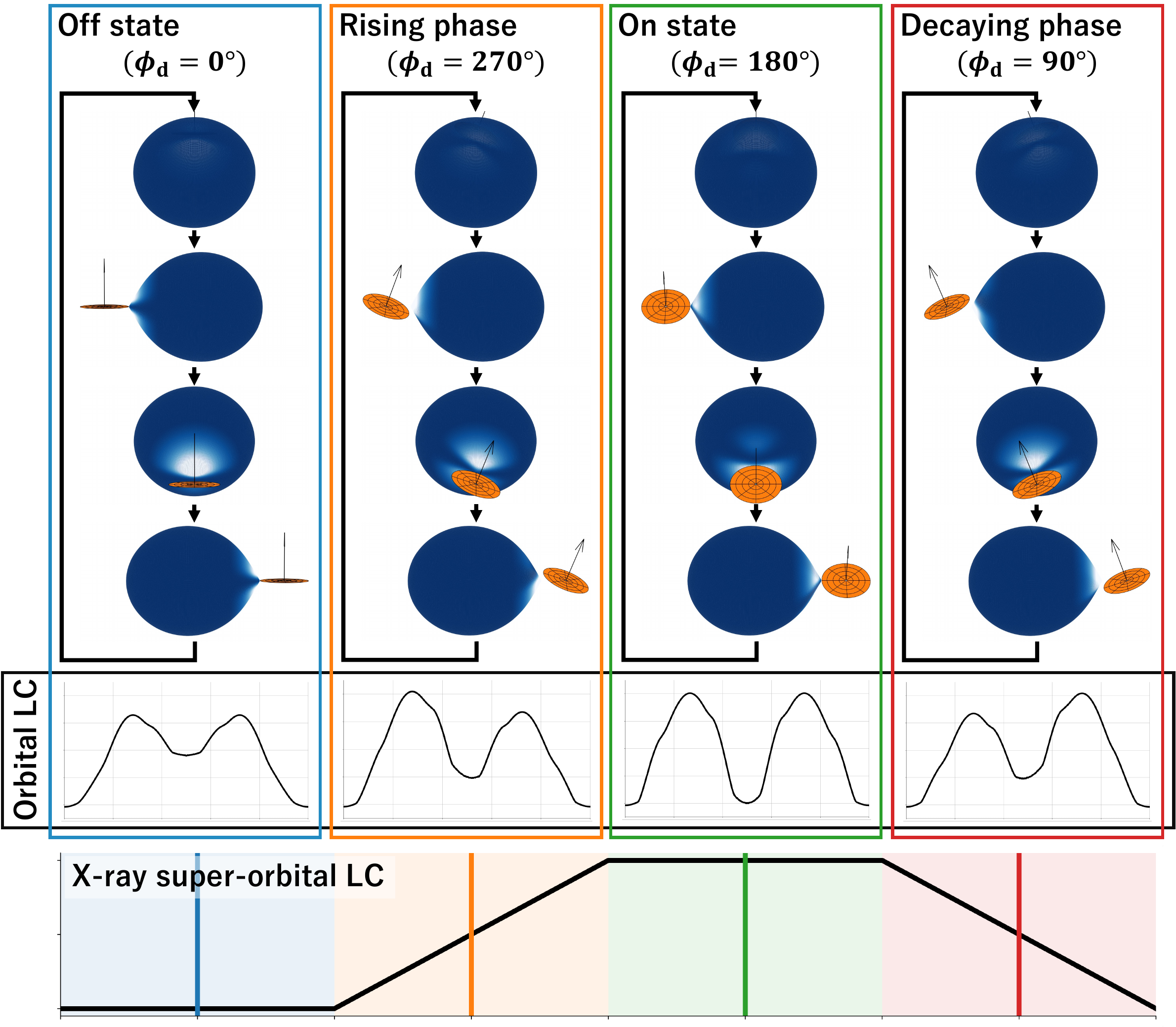}
    \caption{
        A schematic illustration of the relationship between X-ray super-orbital modulation, binary geometry, and orbital light curves in the case of retrograde precession.
        The parameter $\phidisk$ represents the angle of disk precession; since it is defined such that the direction of the orbit is considered positive, $\phidisk$ decreases with time in the case of retrograde precession.
        {
            Alt text: This figure shows the shape of the orbital light curve during the four super-orbital phases (off state, rising phase, on state, and decaying phase) as well as simulated images of the binary as viewed from the observer.
        }
    }
    \label{fig:retrograde}
\end{figure*}

This model also provides an explanation for the asymmetry of the double peak heights during the super-orbital transition phases.
The disk and donor are side by side at each of double peak, and the disk is tilted laterally with respect to the line-of-sight in super-orbital transition phases.
By applying logic similar to that for the inferior conjunction, the peak becomes larger when the disk tilts toward the donor than the opposite (\zcref{fig:disk_tilt2}).
The OUV irradiation onto the disk also contributes to the asymmetry.
The disk side facing the donor is illuminated by OUV radiation from the donor, and heating and scattering on the disk surface makes it brighter as well.
Furthermore, the donor surface visible from the OUV irradiated disk side is heated by X-ray irradiation, making the OUV irradiation stronger.
The lateral tilt of the disk in the super-orbital transition phase reverses depending on whether it is rising or decaying phase, so which of the left and right peaks is higher will also be inverted.
The correspondence between the double peak asymmetry and the super-orbital rising/decaying depends on the precessing direction relative to the orbital rotation.
In the retrograde case, the left (right) peak is higher during the super-orbital rising (decaying) phase; the opposite relationship holds in the prograde case (\zcref{fig:retrograde}).
As can be seen from \zcref{fig:peak-assymetry}, the left peak is higher during the rising phase and vice versa in SMC\,X-1, suggesting its disk precession is retrograde.

Considering the change in the disk inclination relative to the celestial plane and the resulting change in its apparent size, a positive correlation between the optical flux from the disk and the X-ray flux is expected.
This can explain the weak correlation shown in \zcref{fig:opt-X} between the X-ray flux and the average height of the double peak.

\subsubsection{Computational implementation}
\label{sec:model_implement}
We obtained optical fluxes by dividing the donor and disk surfaces into fine surface-elements (SEs), calculating the temperature and blackbody radiation of each SEs, and summing their contribution.
Our model is implemented in Python using NumPy \citep{haris_2020}, SciPy \citep{virtanen_2020}, and Astropy \citep{astropy_2013,astropy_2018,astropy_2022}.
Additionally, the numerical derivation of the Roche-lobe was accelerated with Cython \citep{behnel_2011}, and CuPy (Okuta et.al. 2017)\footnote{Okuta et al. 2017, Machine Learning Systems Workshop at NIPS 2017 \footnoteurl{http://learningsys.org/nips17/assets/papers/paper_16.pdf}} was employed for parallelization of the temperature and flux calculations.

First step is the computation of the geometry.
We neglected the orbital eccentricity and evolutions of $\Porb$ and $\aorb$, given the very small observed values ($e\lesssim10^{-3},\ \Pdot\sim-10^{-6}\>\Year^{-1}$; \citealp{raichur_2010,inam_2010,hu_2019}).
The coordinates in this calculation are normalized by $\aorb$.
Thus the shape of the Roche potential is determined solely by the mass ratio $q$ ($=\Mx/\Ms$; where $\Mx$ and $\Ms$ are pulsar and donor masses, respectively).
Moreover, since the projected semi-major axis of the pulsar orbit ($a_\Xray\sin{\iorb}$) has already been precisely determined ($=53.5769(6)\>\lightsec$; \citealp{raichur_2010}), $\aorb$ can be derived if both $q$ and the orbital inclination $\iorb$ are given.
We made a code deriving the Roche-lobe, based on the assumption that the donor of SMC\,X-1 fills the Roche-lobe.
The dimensionless Roche-potential $\Omega$ is expressed as follows in terms of $(r, \theta, \phi)$ and $q$ \citep{leahy_2015}:
\begin{align}
    \Omega=&q\left(\frac{1}{\sqrt{r^2-2r\sin{\theta}\cos{\phi}+1}}-r\sin{\theta}\cos{\phi}\right)\nonumber\\
    &+\frac{1}{r}+\frac{q+1}{2}r^2\sin^2{\theta}.\label{eq:roche_potential}
\end{align}
First, the $L_1$ point is determined as where $\diff{\Omega}/\diff{r}=0$ with $(\theta,\phi)=(\pi/2,0)$.
Then, we obtained the Roche-lobe by finding the $r$ that yields the same $\Omega$ as that at $L_1$ for various $(\theta, \phi)$.
Since these problems cannot be solved analytically, we used the Brent's method \citep{brent_1973} implemented in SciPy\footnote{\texttt{scipy.optimize.cython\_optimize.brentq}}.
Subsequently, the normal vectors and surface gravity were obtained for each SE by calculating the gradient of $\Omega$.
These calculations were performed in 64 and 128 equal latitude and longitude directions respectively, to obtain 8192 SEs of the Roche-lobe.
Note that for the convenience of parameter exploration, this code is implemented using $\logq$, the common logarithm of $q$, as the input value rather than $q$ itself.
We assumed a thin and circular disk for calculating its radiation and mutual occultation.
This is based on the assumption that the disk shadow that causes X-ray super-orbital modulation is formed only by the innermost region of the disk around the pulsar that is negligibly small compared to the entire disk.
The outermost radius of the disk is parameterized as $\rdisk$, a fraction of the distance from the pulsar to $L_1$, with a maximum value of 1.
The disk was then divided into 32 and 64 sections in the radial and azimuthal direction respectively, for a total of 2048 SEs.
The disk tilt angle $\thetadisk$ is defined as the angle between the normal vectors of the orbital and disk plane.
The phase of the precession is parameterized by $\phidisk$, the counterclockwise rotation angle viewed from above.
We set $\phidisk=0$ when the disk tilts away from the observer, that occurs around the middle of the off state.
Hereafter, subscripts $i$ and $j$ will be used as identifiers for SEs on the star and disk, respectively.

Next step is the calculation of the radiation from each of SEs.
The temperature of the donor surface was derived under the assumption of local thermal equilibrium, taking into account the two effects of gravity-darkening and X-ray irradiation.
The following equation determines the temperature $T_i$ for each SE of the donor:
\begin{equation}
    \sigma T_i^4 = \sigma\Tstar\left(\frac{g_i}{g_0}\right)^\beta+(1-\albedoS)\frac{\eta_i\Lx}{4\pi d_i^2}\cos{\theta_i},
\end{equation}
where $\sigma$ is the Stefan–Boltzmann constant, $\Tstar$ is a surface temperature at the pole of the donor, $g_i$ and $g_0$ are surface gravities at the SE and the donor pole respectively, $\beta$ is a gravity-darkening index, $\albedoS$ is an albedo of the stellar surface, $\Lx$ is an X-ray luminosity of the pulsar, $d_i$ is a distance from the pulsar to the SE, $\eta_i$ represents an effect of X-ray absorption by the disk, and $\theta_i$ is an X-ray incident angle on the SE.
To simplify the model, we assumed the X-ray emission from the pulsar itself (i.e., without the disk shadow) is isotropic.
This may be an oversimplification, but since the pulse period $(=0.71\>\Second)$ is sufficiently shorter than $\Porb\,(=3.89\>\Days)$, it is considered acceptable at least for the azimuthal direction around the spin axis.
The disk shadow is assumed to be mirror-symmetric with respect to the disk plane and axisymmetric around the disk axis.
We used a logistic function as a formulation of $\eta_i$:
\begin{equation}
    \eta_i=\left[1+\exp{\left(-2\frac{|h_i|-\hdisk}{\hbdry}\right)}\right]^{-1},
\end{equation}
where $h_i$ is an elevation angle of the SE from the disk plane as seen from the pulsar, $\hdisk$ is a scale height of the disk shadow, and $\hbdry$ is a width of the shadow boundary region.
This function was selected to represent the continuous transition characteristic of the X-ray super-orbital modulation phenomenologically.
Note that our disk modeling (thin and circular shape, highly symmetric shadow) may also be oversimplified, since the X-ray super-orbital modulation of SMC\,X-1 is usually explained with a ``warped'' misaligned disk.
Then the blackbody spectrum with temperature $T_i$ is integrated while weighted by the \satellite{TESS} response function to obtain the surface brightness of the SE, $I_i$:
\begin{equation}
    I_i = \frac{\int R(\nu)B(\nu,T_i)\,\diff{\nu}}{\int R(\nu)\,\diff{\nu}},
    \label{eq:star_brightness}
\end{equation}
where $R$ is a response function of the \satellite{TESS} instrument and $B$ is the Planck distribution function.
The \satellite{TESS} response function is obtained by unit conversion and linear interpolation on the transmittance table downloaded from NASA \satellite{TESS} Science Support Center\footnote{\footnoteurl{https://heasarc.gsfc.nasa.gov/docs/tess/documentation.html}}.
The disk temperature is determined by first applying uniform heating across the entire disk, then further heating it with OUV irradiation from the donor, and considering these heating to be in equilibrium with cooling via blackbody radiation.
However, calculating irradiation for all combinations of donor and disk SEs would be computationally very intensive.
Thus we made the approximation that the entire disk is uniformly illuminated with the same amount of OUV flux at the pulsar position ($F_{(\Star\rightarrow\Pulsar)}$).
This approximation reduced the computation time to about 1/500.
The justification of the approximation is described in \zcref{sec:test-singleT}.
The temperature of each SE on the disk is derived by the following equation:
\begin{align}
    \sigma T_j^4=&\sigma\Tdisk^4+(1-\albedoD)F_{(\Star\rightarrow\Pulsar)},
\end{align}
where $\Tdisk$ is a base temperature of the disk and $\albedoD$ is an albedo of the disk surface.
The subsequent calculation of brightness by the blackbody radiation $I_{\mathrm{bb},j}$ is the same as for the donor:
\begin{equation}
    I_{\mathrm{bb},j} = \frac{\int R(\nu)B(\nu,T_j)\,\diff{\nu}}{\int R(\nu)\,\diff{\nu}}.
\end{equation}
The unabsorbed portion of the OUV irradiation is assumed to be isotropically scattered by the Thomson process on the disk surface, and the brightness by this scattering is then denoted as $I_{\mathrm{scat},j}$.
The resulting surface brightness of the disk $I_j$ is the sum of these two components, blackbody radiation and Thomson scattering:
\begin{equation}
    I_j = I_{\mathrm{bb},j} + I_{\mathrm{scat},j}.
\end{equation}

Finally, the flux is obtained by adding up the brightness of each SE while taking into account the mutual occultation of the donor and the disk.
The following is the calculation of the donor flux $f_\Star$:
\begin{equation}
    f_\Star=\frac{1}{d^2}\sum_i\delta_i\left[1 + u(1-\cos{\Theta_i})\right]I_iA_i\cos{\Theta_i},\label{eq:f_star}
\end{equation}
where $d$ is a distance to SMC\,X-1, $\delta_i$ is a visibility factor of the SE (1 when visible, 0 when not), $u$ is a limb-darkening index, $\Theta_i$ is an angle between line of sight and normal vector of the SE, and $A_i$ is the area of the SE.
For the disk, it is almost same as for the donor, except that the limb-darkening is not considered:
\begin{equation}
    f_\Disk=\frac{1}{d^2}\sum_j\delta_jI_jA_j\cos{\Theta_j},
\end{equation}
where the definitions of the symbols are similar to those in \zcref{eq:f_star}.
The sum of them is the optical flux $f_\Opt$ output by the model:
\begin{equation}
    f_\Opt=f_\Star+f_\Disk.
\end{equation}
We assumed the interstellar and intergalactic dust extinction to be negligible.

Additionally, this model can predict the shape of X-ray light curve.
The value of $\eta$ calculated for the direction toward the observer ($\eta_\mathrm{obs}$) represents the fraction of X-rays in that direction not absorbed by the disk, allowing it to be used to predict super-orbital X-ray variations.
The simplest method to obtain the X-ray flux $f_\Xray$ from $\eta_\mathrm{obs}$ is to multiply by an unobscured X-ray flux $\Fx$:
\begin{equation}
    f_\Xray=\delta_\Pulsar\eta_\mathrm{obs}\Fx,
\end{equation}
where $\delta_\Pulsar$ is a visibility factor of the pulsar (0 during the eclipse, otherwise 1).

\subsection{Light curve fitting}
We fitted optical and X-ray light curves of SMC\,X-1 with our modified ellipsoidal modulation model, to test the model and to estimate binary parameters.
To confirm the effect of super-orbital modulation, it is desirable to use light curves covering the whole super-orbital period.
Sectors 1 and 2 form an almost continuous light curve covering the super-orbital period due to the two sectors being adjacent, and the behavior of X-ray super-orbital modulation is simpler than that in sectors 27 and 28 (\zcref{fig:xray_superorbital}).
Therefore, we decided to use the light curves in cycles 1408--1421, which are contained within sectors 1 and 2.

\subsubsection{Configuration of the MCMC sampling}
\label{sec:mcmc_sampling}
We employed a Markov chain Monte Carlo (MCMC) method to search for a parameter set capable of reproducing optical and X-ray light curves with our model.
We utilized ``emcee'' \citep{foreman_2013}, a Python-implemented affine-invariant MCMC ensemble sampler of \citet{goodman_2010}.
We assumed that the disk precesses at a constant angular velocity $\phidot$ during cycles 1408--1421, with an initial precession phase $\phizero$.
The reference point of the disk precession (i.e., the point where $\phidisk=\phizero$) was set to $\PhiOrb=1409.0$.
This choice is motivated by the fact that cycle 1409 marks the reversal of the double-peak asymmetry and corresponds to a characteristic super-orbital phase ($\phidisk\sim180^\circ$).
\zcref[S]{tab:fixedparams,tab:exploredparams} list the fixed and free parameters.
The $\beta$ value was set to 1.0, which is commonly adopted for radiative envelopes.
The $u$ value was obtained by applying spline interpolation to the table of \citet{claret_2017} at $\log{g}=4.2$ \citep{gaia_2023}, metallicity $Z=-1$ \citep{choudhury_2018}, and micro-turbulent velocity $\xi=2\>\kMpS$.
In our analysis, $\logq$ was not treated as a free parameter.
Instead, we sampled 100 fixed values of $\logq$ from a Gaussian distribution based on the observed RVs of the pulsar and donor \citep{raichur_2010, vdm_2007}, and performed an independent MCMC sampling for each value.
The combined posterior samples from all runs define the final parameter distribution.
When $\logq$ was also treated as a free parameter, its value is not uniquely determined and converges to different values within the range $-1.4$ to $-1.0$ with each sampling attempt.
Therefore, it was necessary to impose a constraint on $\logq$.
As a separate case, we repeated the same procedures using a $\logq$ distribution shifted by $+0.08$ (i.e., increase $q$ by 20\%).
The motivation for this additional analysis stems from the fact that the observed donor RV may be underestimated, as discussed later.

\begin{table}
    \tbl{
        Fixed parameters
    }{
        \begin{tabular}{cccc}
            \hline
            Param. & Value & Description & Ref.\\
            \hline
            $d$ & 62\>\kpc & Distance to SMC\,X-1 & [1]\\
            $\albedoS$ & 0.3 & Albedo of the donor & [2]\\
            $\albedoD$ & 0.3 & Albedo of the disk & [2]\\
            $\beta$ & $1.0^*$ & Gravity-darkening index & -\\
            $u$ & 0.22 & Limb-darkening index & [3]\\
            \multirow{2}{*}{$\logq^\dagger$} & $-1.17(3)$ & \multirow{2}{*}{Logarithmic mass-ratio} & [4], [5]\\
            & $-1.09(3)^\ddagger$ & & -\\
            \hline
        \end{tabular}
    }
    \label{tab:fixedparams}
    \begin{tabnote}
        [1] \citet{graczyk_2020}; [2] \citet{heemsherk_1989}; [3] \citet{claret_2017}; [4] \citet{raichur_2010}; [5] \citet{vdm_2007}\\
        \footnotemark[$*$] A typical value used for radiative envelopes\\
        \footnotemark[$\dagger$] The bracketed number is the 1$\sigma$ uncertainty of the last digit.\\
        \footnotemark[$\ddagger$] Increased $q$ by 20\% from the observed RV ratio.
    \end{tabnote}
\end{table}

\begin{table}
    \tbl{
        Free parameters
    }{
        \begin{tabular}{cc}
            \hline
            Param. & Description\\
            \hline
            $\iorb$ & Orbital inclination\\
            $\Tstar$ & Temperature at the donor pole\\
            $\Tdisk$ & Base temperature of the disk\\
            $\Lx$ & X-ray luminosity of the pulsar\\
            $\rdisk$ & Normalized disk outermost radius\\
            $\thetadisk$ & Disk tilt angle\\
            $\hdisk$ & Height of the disk shadow\\
            $\hbdry$ & Boundary width of the disk shadow\\
            $\phizero$ & Initial disk precession phase\\
            $\phidot$ & Disk precession rate\\
            $\Fx$ & Unobscured X-ray flux\\
            \hline
        \end{tabular}
    }
    \label{tab:exploredparams}
    \begin{tabnote}
    %\footnotemark[$*$] Cadences of FFIs
    \end{tabnote}
\end{table}

\subsubsection{Results of the MCMC sampling and discussion}
\label{sec:mcmc_result}
\begin{figure*}
    \includegraphics[width=\linewidth]{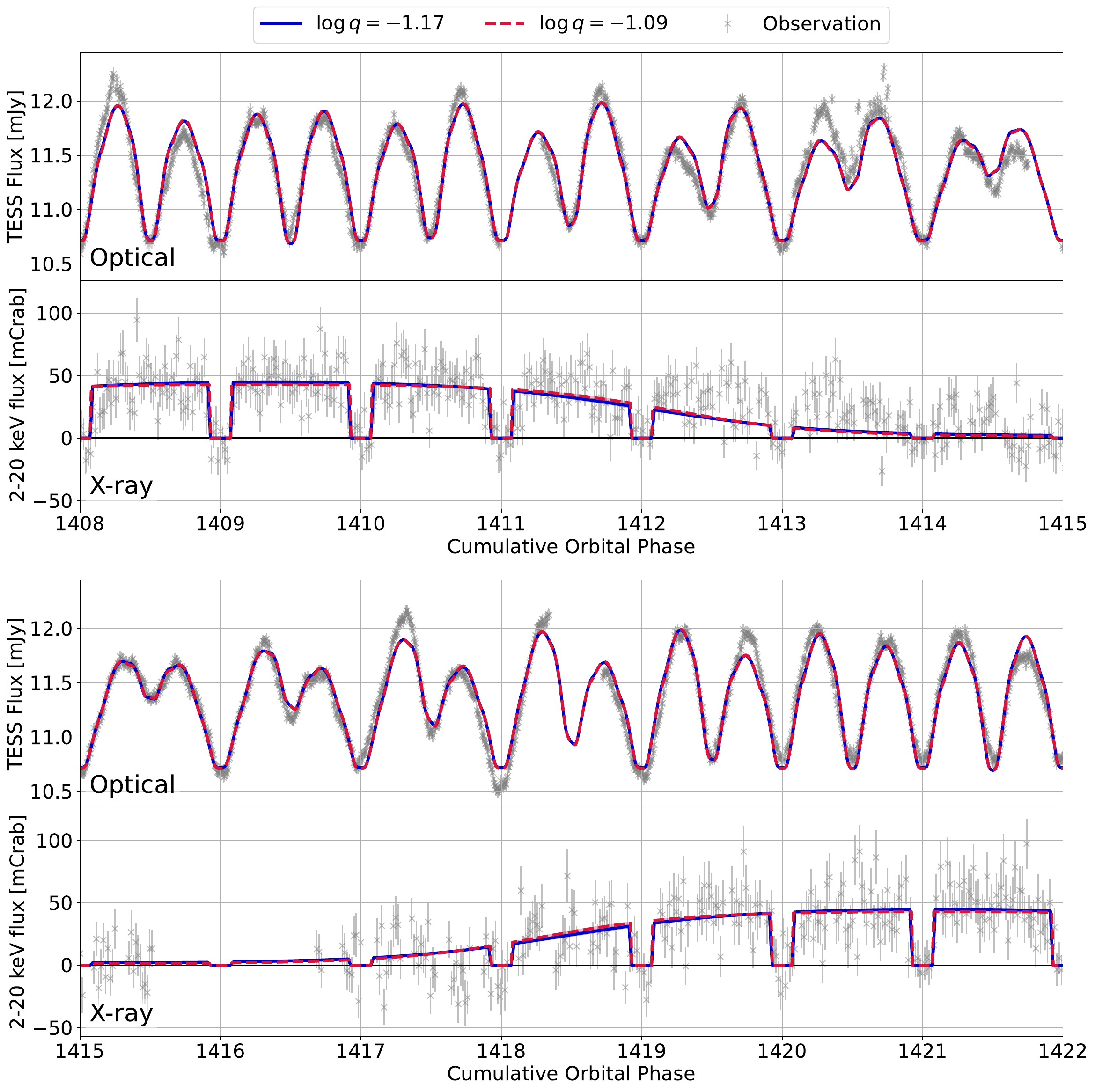}
    \caption{
        Best-fit optical and X-ray model light curves of two $\logq$ distributions.
        The 1$\sigma$ ranges are shown as shaded areas.
        {
            Alt text: Two adjacent pairs of panels are arranged vertically, totaling four panels.
            The upper panel in each pair shows the observed and two best-fit optical light curves, while the lower one shows the same for X-rays.
            The upper and lower pairs are light curves from cycle 1408 to 1414 and from cycle 1415 to 1421, respectively.
        }
    }
    \label{fig:lc_and_models}
\end{figure*}

\begin{figure*}
    \includegraphics[width=\linewidth]{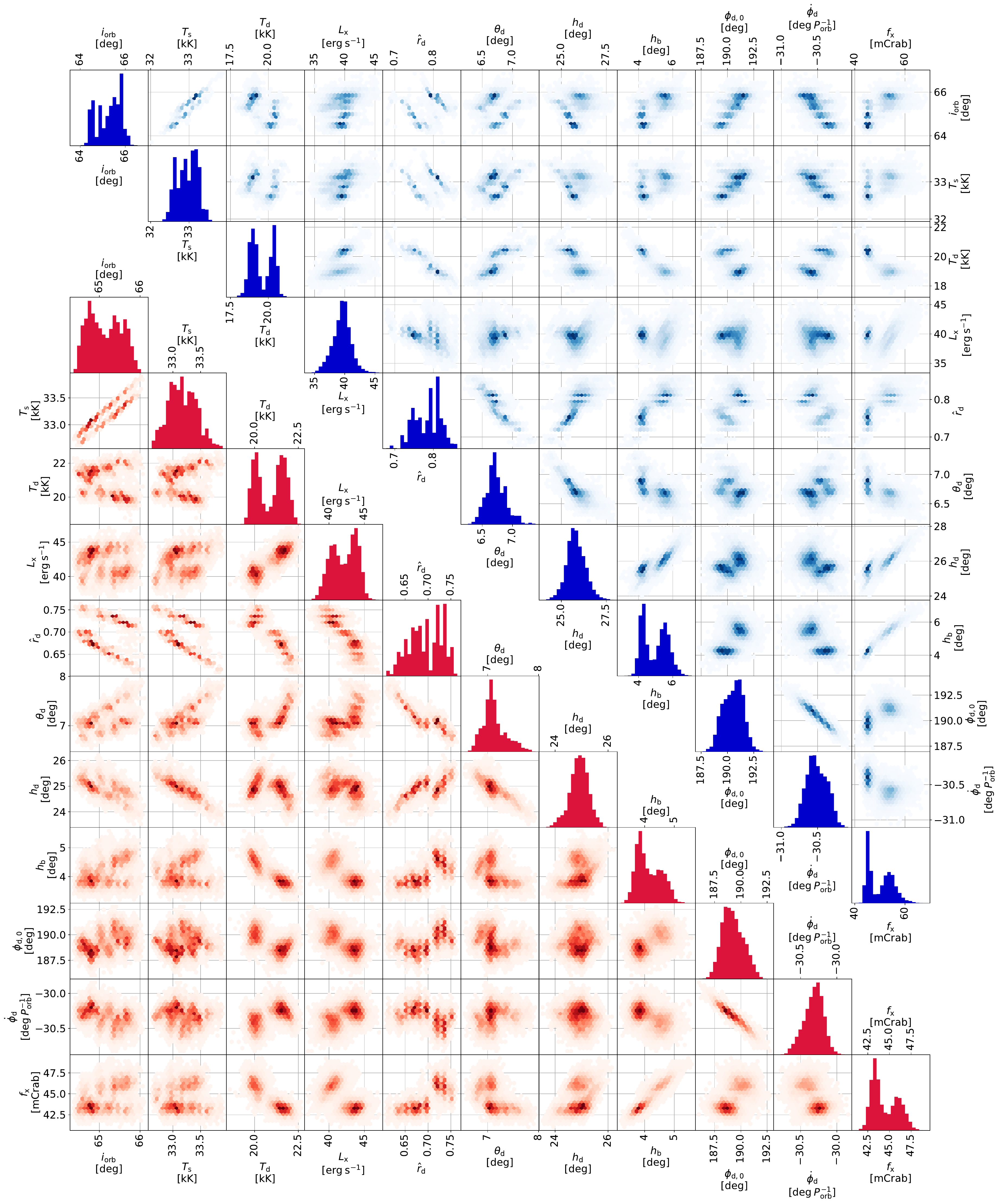}
    \caption{
        Corner plots of posterior parameter distributions for both $\log{q}=-1.17, -1.09$ cases, shown in the upper right and lower left, respectively.
        A linear color scale is used for the two-dimensional histograms in the off-diagonal panels.
        {
            Alt text: Corner plots of two posterior parameter distributions obtained by MCMC sampling.
        }
    }
    \label{fig:corner}
\end{figure*}

We present the best-fit model light curves in \zcref{fig:lc_and_models}.
The parameter distributions and derived binary properties inferred from the MCMC analysis are summarized in \zcref{tab:mcmc_results}, while the corresponding posterior distributions are visualized as corner plots in \zcref{fig:corner}.
Our model succeeded in roughly reproducing the observed optical and X-ray light curves.
This result provides ``optical'' observational evidence supporting the validity of the precessing disk scenario for super-orbital modulation in SMC\,X-1.
The estimated pulsar mass $\Mx=1.11\pm0.07\>\MO$ is consistent with the previous studies \citep{vdm_2007,rawls_2011,coe_2013}, when $\logq=-1.17$.
The negative value of $\phidot$ implies the retrograde precession, as predicted by the qualitative discussion in \zcref{sec:model_concept}.
The retrograde precession of the misaligned disk can be understood as a natural consequence of tidal torque exerted by the donor.
Furthermore, viscous torque and radiation torque also naturally cause retrograde precession \citep{ogilvie_2001}.
Nearly identical light curves were generated for both $\logq=-1.17, -1.09$, and this is consistent with that the value of $\logq$ was not uniquely determined when treating it as a free parameter.

\begin{table}
    \tbl{
        Average model parameters and $1\sigma$ uncertainties obtained from MCMC sampling (top 11 rows), and binary quantities derived from them (bottom 5 rows).
    }{
        \begin{tabular}{cccc}
            \hline
            Param. & $\logq=-1.17$ & $\logq=-1.09$ & Unit\\
            (${\chi^2_\text{avg}}^*$) & ($4.9\times10^{4}$) & ($4.9\times10^4$) & \\            \hline
            $\iorb$ & $65.2\pm0.5$ & $65.1\pm0.4$ & \Deg\\
            $\Tstar$ & $33.0\pm0.3$ & $33.2\pm0.2$ & k\Kelvin\\
            $\Tdisk$ & $19.6\pm0.7$ & $20.9\pm0.8$ & k\Kelvin\\
            $\Lx$ & $39\pm2$ & $42\pm2$ & $10^{38}\>\EpS$\\
            $\rdisk$ & $0.78\pm0.03$ & $0.70\pm0.04$ & -\\
            $\thetadisk$ & $6.7\pm0.2$ & $7.1\pm0.2$ & \Deg\\
            $\hdisk$ & $25.8\pm0.5$ & $24.9\pm0.4$ & \Deg\\
            $\hbdry$ & $4.9\pm0.7$ & $4.1\pm0.4$ & \Deg\\
            $\phizero$ & $190.6\pm0.9$ & $189.2\pm1.0$ & \Deg\\
            $\phidot$ & $-30.5\pm0.1$ & $-30.3\pm0.1$ & $\Deg\>\Porb^{-1}$\\
            $\Fx$ & $50\pm5$ & $44\pm1$ & m\Crab\\
            \hline
            $\aorb$ & $27.1\pm0.2$ & $27.5\pm0.1$ & $\RO$\\
            $\tilde{R}_\Star$$^\dagger$ & $16.55\pm0.08$ & $16.42\pm0.08$ & $\RO$\\
            $r_\Disk$$^\ddagger$ & $5.4\pm0.1$ & $5.0\pm0.1$ & $\RO$\\
            $\Ms$ & $16.6\pm0.3$ & $17.0\pm0.2$ & $\MO$\\
            $\Mx$ & $1.11\pm0.07$ & $1.35\pm0.10$ & $\MO$\\
            \hline
        \end{tabular}
    }
    \label{tab:mcmc_results}
    \begin{tabnote}
        \footnotemark[$*$] Average $\chi^2$ of sampled parameter sets. ($\text{d.o.f}=3064$)\\
        \footnotemark[$\dagger$] A radius of a sphere with the same volume as the Roche lobe\\
        \footnotemark[$\ddagger$] A disk outermost radius
    \end{tabnote}
\end{table}

\begin{figure}
    \includegraphics[width=\linewidth]{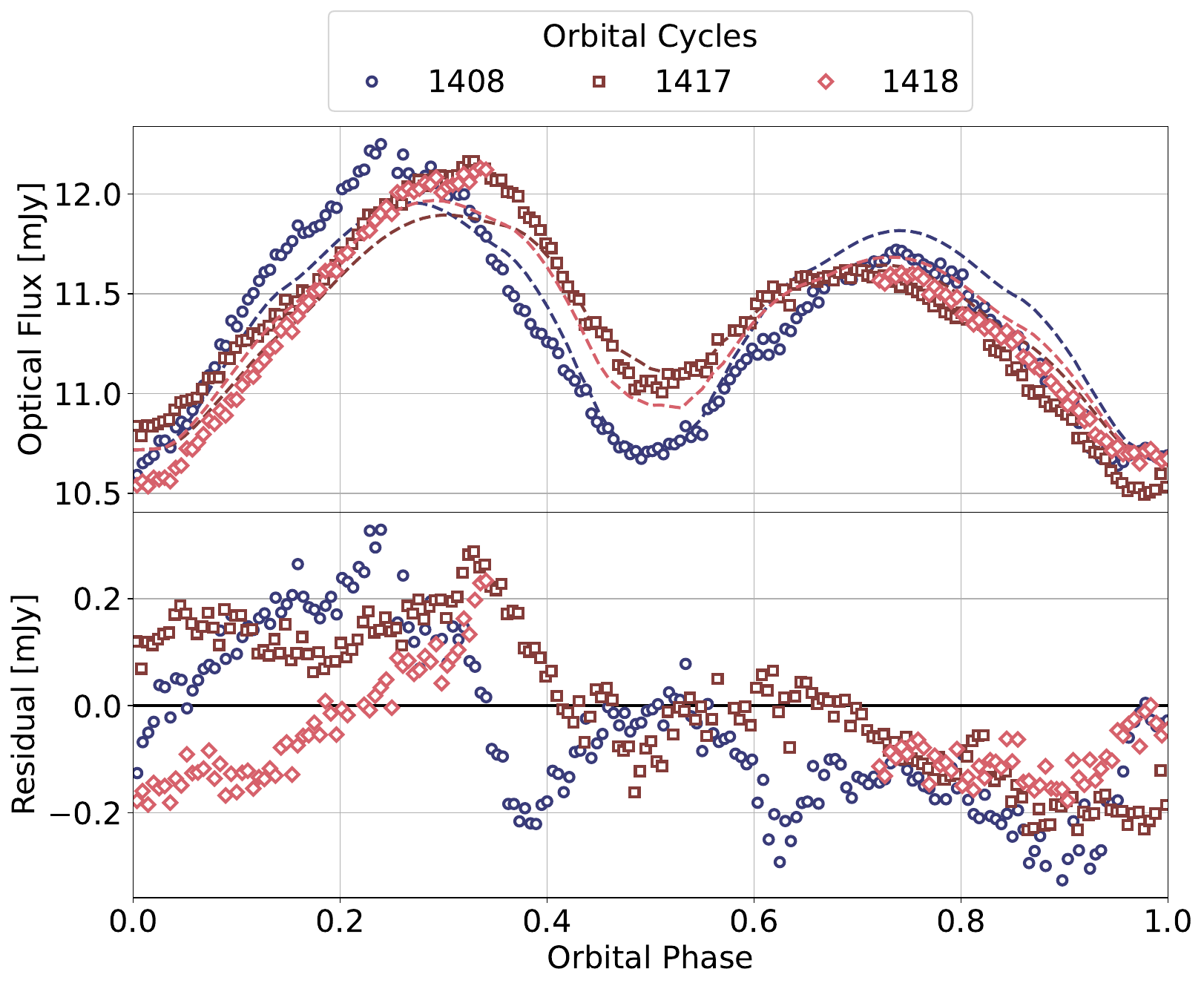}
    \includegraphics[width=\linewidth]{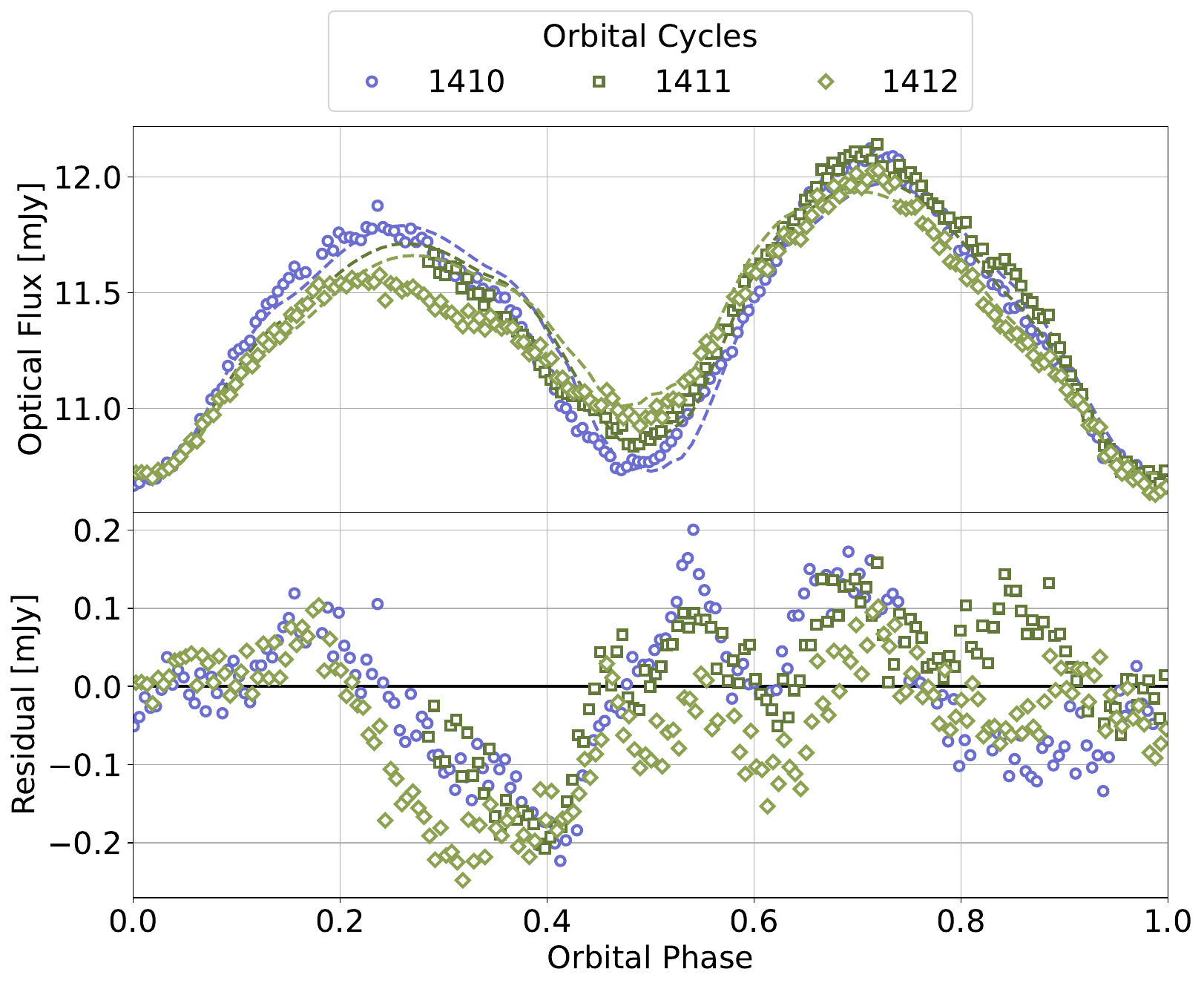}
    \caption{
        Comparison of residuals in orbital cycles with similar light curve shapes.
        Dashed lines are the model light curves in each orbital cycles.
        {
            Alt text: Two adjacent pairs of panels are arranged vertically, totaling four panels.
            The upper panel in each pair shows the observed and model optical light curves, while the lower one shows residuals.
            The upper and lower pairs respectively show data for cycles 1408, 1417, and 1418, and cycles 1410 through 1412.
        }
    }
    \label{fig:residuals}
\end{figure}

The model and observations are generally consistent, but there are significant discrepancies locally.
In particular, for cycles with asymmetric optical double-peak heights, the asymmetry in the model is systematically smaller than in the observations.
Additionally, the discrepancy between the model and observations is remarkably prominent for cycles 1413 and 1414.
Then, we performed extra MCMC sampling individually for the orbital light curves each of cycles 1413 and 1414, while treating $\thetadisk$, $\phidisk$, and $\hdisk$ as free parameters and fixing the others to the best-fit parameter set of the $\logq=-1.17$ case in \zcref{tab:mcmc_results}.
Note that $\phidisk$ is assumed to be constant over the orbital cycle in this MCMC sampling.
\zcref[S]{fig:cutlc,tab:cut_mcmc} are the resulting light curves and the parameter distributions, respectively.
Compared to fitting the entire light curve for sectors 1 and 2, the fit is significantly improved, suggesting that there are shifts from average trend in precession phase, the disk tilt angle or thickness during these two cycles.
We show residuals in orbital cycles with similar light curve shapes in \zcref{fig:residuals}.
Adjacent cycles show similar residual trends.
However, the residuals for cycles with high left peaks (top panel in \zcref{fig:residuals}) and those with high right peaks (bottom panel) do not appear to match each other when reversed left-to-right.
These results may suggest that the disk shadow has a non-axisymmetric structure and that it varies over several cycles.
Such behavior is qualitatively consistent with theoretical predictions that, under high accretion rates, strong radiation torques can cause the disk to self-warp \citep{pringle_1996,ogilvie_2001} and form a geometrically thick accretion flow \citep{ohsuga_2011}.
Therefore, systematic deviations of our model from observations may be a manifestation of these disk structures.
Another possible explanation for the model-observation discrepancy is variations in intrinsic X-ray luminosity.
In our modeling, we assumed that the long-term X-ray variability is solely due to occultation by the precessing disk and that the intrinsic X-ray luminosity remains constant over time.
While this assumption successfully reproduces the overall long-term X-ray light curve, some local discrepancies remain at several epochs, some of which appear to coincide with increases or decreases in optical flux (e.g., $\PhiOrb\approx1412.25, 1413.25, 1414.75, 1417.25$).
This can be interpreted as that variations in X-ray luminosity caused the optical flux variations through irradiation of the donor.

\begin{figure}
    \includegraphics[width=\linewidth]{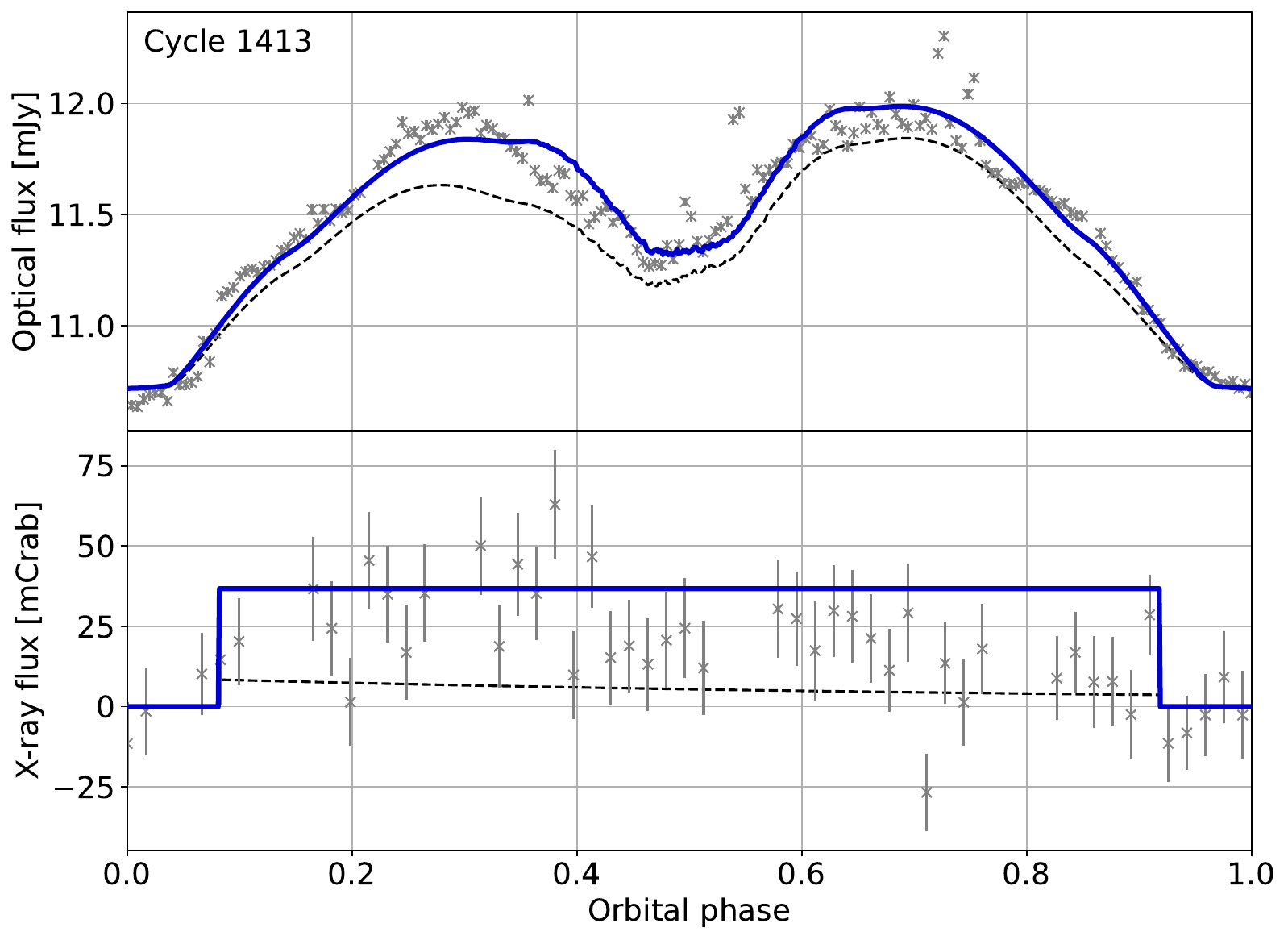}
    \includegraphics[width=\linewidth]{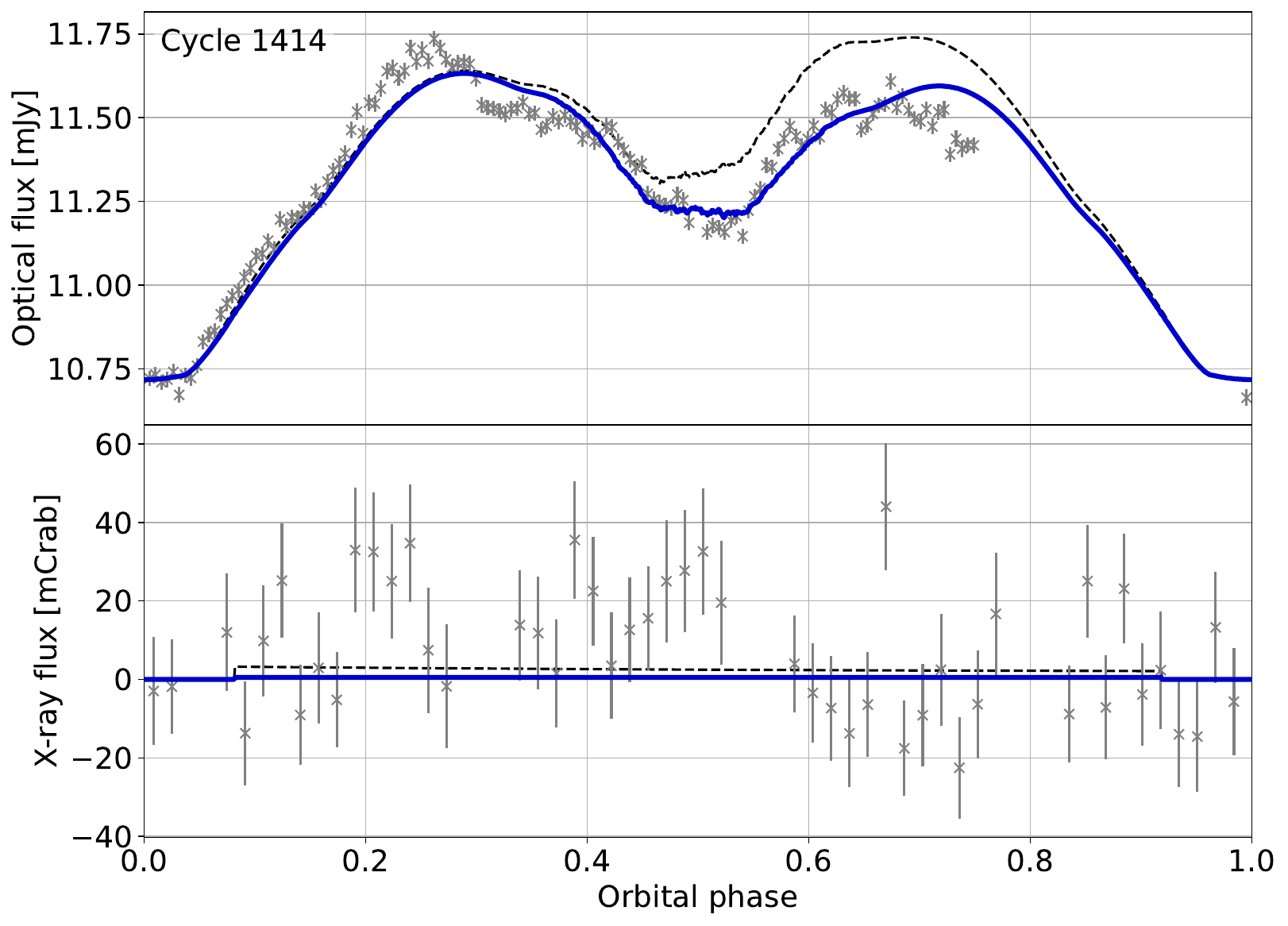}
    \caption{
        Model light curves in cycles 1413 and 1414 obtained by individual MCMC sampling.
        The blue solid and black dashed lines are the light curves from the individual fitting, and from the overall fitting of sectors 1 and 2 (\zcref{fig:lc_and_models}), respectively.
        {
            Alt text: Two adjacent pairs of panels are arranged vertically, totaling four panels.
            The upper panel in each pair shows the observed and model optical light curves, while the lower one shows the same for X-rays.
            Each pair corresponds to the two cycles.
        }
    }
    \label{fig:cutlc}
\end{figure}

\begin{table}
    \tbl{
        Parameter distributions in cycles 1413 and 1414 obtained by individual MCMC sampling
    }{
        \begin{tabular}{cccc}
            \hline
            Parameter & Cycle 1413 & Cycle 1414 & Unit\\
            \hline
            $\thetadisk$ & $3.015\pm0.006$ & $7.477\pm0.009$ & deg\\
            $\phidisk$ & $73.7\pm0.1$ & $352.1\pm0.2$ & deg\\
            $\hdisk$ & $21.33\pm0.02$ & $28.57\pm0.02$ & deg\\
            \hline
            $\overline{\phi}_\Disk$$^*$ & 51.5 & 21.2 & deg\\
            \hline
        \end{tabular}
    }
    \label{tab:cut_mcmc}
    \begin{tabnote}
        \footnotemark[$*$] The average precession phase in the given orbital cycle, predicted with the parameter set of $\logq=-1.17$ case in \zcref{tab:mcmc_results}.
    \end{tabnote}
\end{table}

\begin{figure}
    \includegraphics[width=\linewidth]{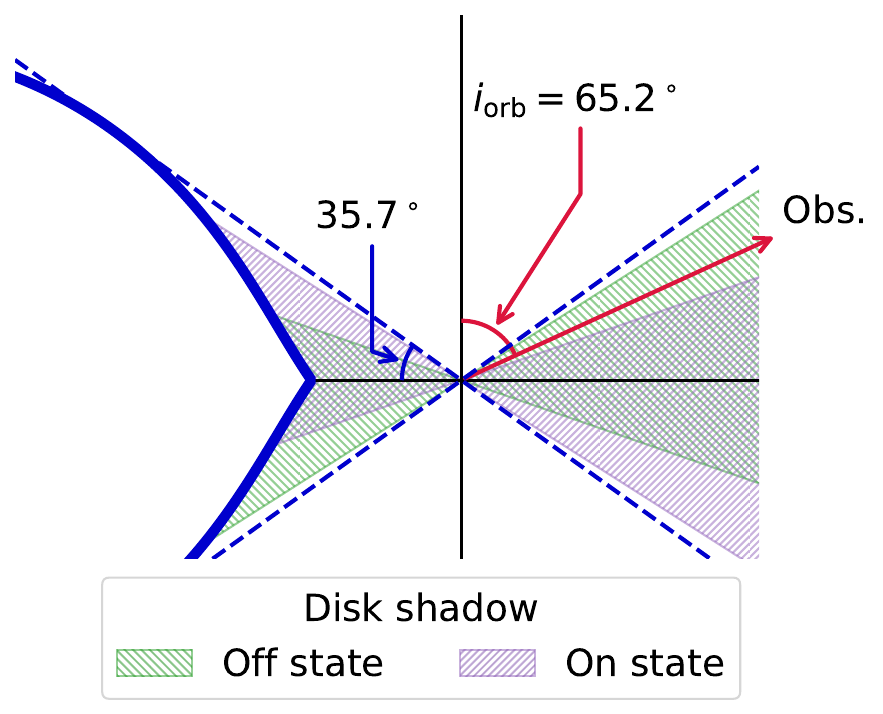}
    \caption{
        Schematic cross-sectional view of the binary in the $X$-$Z$ plane at the inferior conjunction phase ($\phi_\Orb=0.5$), with the parameter set of $\logq=-1.17$ case (cf. \zcref{tab:mcmc_results}).
        The horizontal and vertical solid black lines represent the $X$ and $Z$-axis, respectively, with the $Z$-axis drawn at the position of the pulsar.
        The teardrop-shaped object on the left is the donor, and ``Obs.'' indicates the direction toward the observer.
        Blue dashed lines show the range where X-rays emitted from the pulsar strike the donor, while a red solid line is the line of sight vector.
        The shaded areas are the disk shadow in the off and on states (i.e., $\phi_\Disk=0^\circ, 180^\circ$, respectively).
        The disk itself is not drawn.
        {
            Alt text: This compares the geometry of the X-rays irradiating the donor with that of the X-rays traveling toward the observer, at the inferior conjunction.
        }
    }
    \label{fig:xz-plane}
\end{figure}

Focusing on the parameter distribution, we find that $\Lx$ is extremely large, $\approx4\times10^{39}\>\EpS$, which is about an order of magnitude higher than typical values inferred in previous X-ray observations ($\approx5\times10^{38}\>\EpS$; e.g., \citealp{bonnet_1981,pike_2019}).
\citet{pradhan_2020} derived a broadband (1--70\,keV) X-ray flux of SMC\,X-1 to be $\approx3\>\EpSC$.
Assuming a distance of 62 kpc, this corresponds to a luminosity of $\approx1.4\times10^{39}\>\EpS$, which is closer to our value, although our estimate is still higher by a factor of $\approx3$.
In our modeling, $\Lx$ is introduced purely as a parameter controlling the irradiation intensity onto the donor.
Thus, this discrepancy indicates that the X-ray luminosity estimated from observations is insufficient to reproduce the optical super-orbital modulation, even after accounting for additional optical contributions (i.e., OUV reprocessing on the disk).
The most straightforward explanation for this difference is an anisotropy of X-ray emission.
If the X-rays are strongly beamed toward the donor while only a smaller fraction is emitted toward the observer, the required high $\Lx$ could be reconciled with the observed X-ray luminosity.
However, there is no significant difference in the elevation angles from the orbital plane between the vectors to the observer and to the irradiated donor surface according to the obtained parameter distribution (\zcref{fig:xz-plane}).
Therefore, if the spin axis of the pulsar is aligned with the orbital axis, anisotropic X-ray emission alone cannot easily account for this problem.
There are two possible interpretations for the discrepancy in between the observed and our predicted luminosity: a very narrow beam or a spin-orbit misalignment.
In the first one, the beam with a few degree width is emitted at an elevation angle slightly lower or higher than that of the observer; while the center of the beam strikes the donor, we observe the edge of the beam.
For considering the second one, the spin-orbit misalignment, the evolutionary history of SMC\,X-1 is important.
It seems natural to consider that SMC\,X-1 experienced a common envelope (CE) phase, based on its very close and circular orbit ($\aorb\sim0.1\>\AU$, $e\lesssim10^{-3}$).
If sufficient angular momentum exchange between the pulsar and the envelope was carried out during the CE phase, any initial spin-orbit misalignment (e.g., due to a natal kick) would be erased, leading to alignment between the spin and orbital axes.
In other words, the spin-orbit misalignment in such close X-ray binaries could serve as evidence for inefficient mass accretion onto the neutron star during the CE phase.
Earlier theoretical studies suggested that mass accretion onto compact objects within the CE is very massive, because the radiation is trapped in the adiabatic inflow and radiation pressure cannot halt accretion \citep{chevalier_1993,brown_1995,bethe_1998}.
In contrast, recent three-dimensional fluid simulations have revealed that mass accretion is far less efficient than previously thought, as the accretion flow carries angular momentum due to the density gradient of the CE and the orbital motion of the compact object, making it difficult to accrete onto the compact object \citep{macleod_2015,murguia_2017,murguia_2020}.
Therefore, our result favors these recent theoretical studies.
An alternative evolutionary scenario has been proposed by \citet{li_2014},
who argue that SMC\,X-1 did not undergo a common-envelope (CE) phase.
They instead suggest that the orbit was tightened through mass loss via the $L_2$ point or through interaction with a circum binary disk.
In such a scenario, a spin-orbit misalignment can also be naturally expected.
Their argument is based on that the pulsar of SMC\,X-1 should be produced by an electron-capture supernova because of its estimated mass $\Mx\approx1\>\MO$.
However, as will be discussed later, that mass may be underestimated.

Given that the donor is subjected to such intense irradiation, it may exert a non-negligible influence on the observed properties of the donor.
Among them, we believe the RV warrants particular attention.
Generally, the Doppler shifts of emission or absorption lines are believed to represent the RV at the stellar centroid.
However, in the case of SMC\,X-1, the intense X-ray irradiation enhances optical emission from the donor surface facing the pulsar, potentially leading to an underestimation of the RV.
In our model, the apparent RV can be estimated by averaging the RV of all SEs, weighted by their contributions to the observed optical flux.
\zcref[S]{fig:effective_RV} compares the apparent and actual RV.
The estimated apparent RV curve is not a simple sine wave; rather, it has a shape that appears to be clipped at approximately 80\% of its amplitude.
In other words, the observed RV amplitude may be underestimated by about 20\%.
\citet{vdm_2007}, which has been widely cited as the basis for the donor RV of SMC\,X-1, did not ignore X-ray irradiation effects and attempted to mitigate them by excluding absorption lines with large equivalent-width (EW) variations, under the assumption that such lines originate predominantly from the irradiated surface.
However, EW is the relative line strength to the spectral continuum, and since irradiation enhances the continuum as well as the absorption lines, the EW variation may be an insufficient diagnostic of irradiation effects.
In \zcref[S]{fig:vdm_2007}, we compare a simple sine curve and an 80\%-clipped sine curve with the observed RV curve reported by \citet{vdm_2007}, where the clipped sine curve is introduced as a simplified representation of the apparent RV curve predicted by our model.
The simple sine model adopts an amplitude of $20.2\>\kMpS$ following \citet{vdm_2007}, while the clipped sine model uses an amplitude 1.2 times larger, with the waveform clipped at 80\% of its peak value.
As shown in the figure, both curves reproduce the observational data with comparable fidelity. Therefore, our prediction of a $\approx20\%$ suppression in the observed RV amplitude is consistent with the observations of \citet{vdm_2007}.
At the same time, their observational data do not allow us to determine which of the two models is more plausible, and future high-precision spectroscopic observations are needed to obtain a more precise RV curve that can distinguish between the two scenarios.
Estimates of pulsar mass are generally based on the mass ratio $q$ derived from the donor and pulsar RV ratio.
Therefore, the pulsar masses may also be underestimated.
When performing MCMC sampling with increasing $q$ by 20\% (i.e., $\logq=-1.09$ case in \zcref{tab:mcmc_results}), the resulting pulsar mass becomes $\Mx=1.35\>\MO$, a more common value for NSs.

\begin{figure}
    \includegraphics[width=\linewidth]{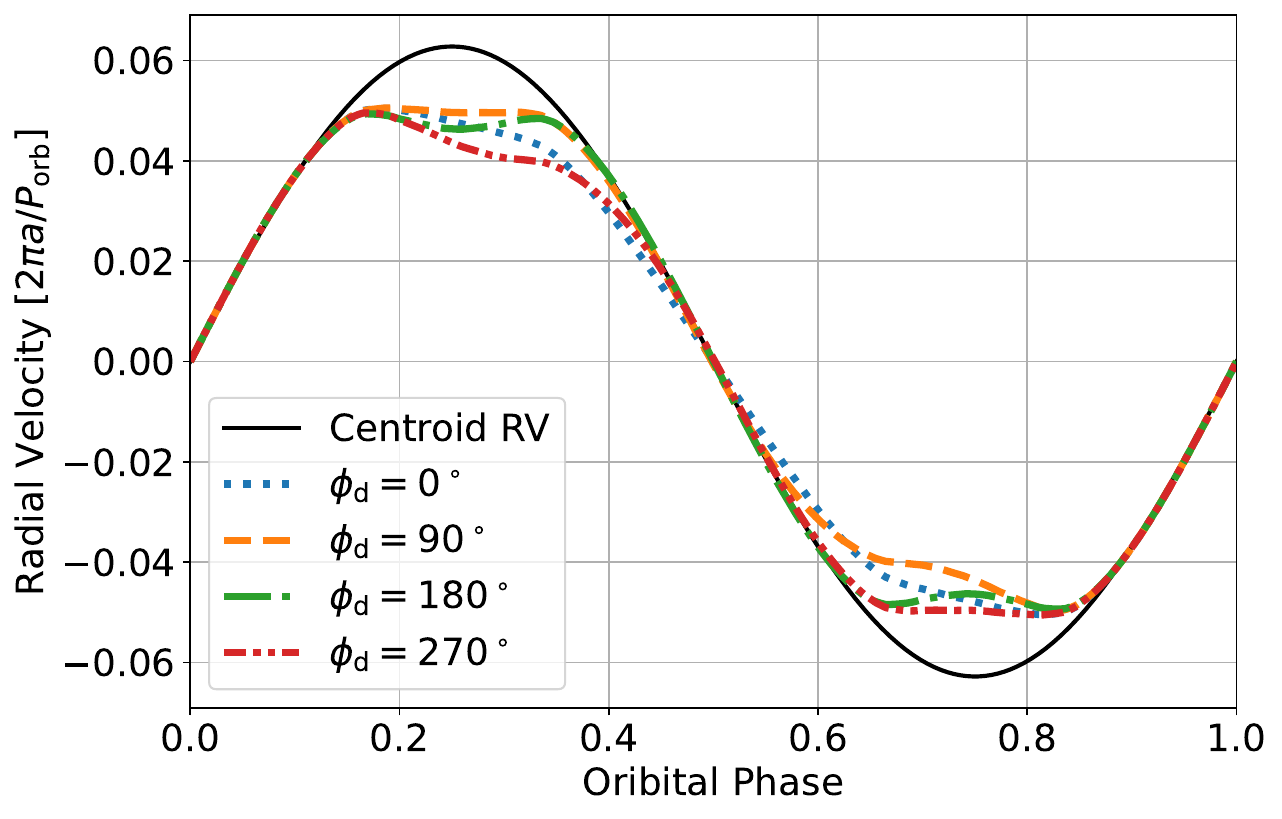}
    \caption{
        Apparent and actual radial velocities of the donor including irradiation effects, in several super-orbital phases.
        {
            Alt text: The vertical and horizontal axes show radial velocity and orbital phase, respectively.
        }
    }
    \label{fig:effective_RV}
\end{figure}

\begin{figure}
    \includegraphics[width=\linewidth]{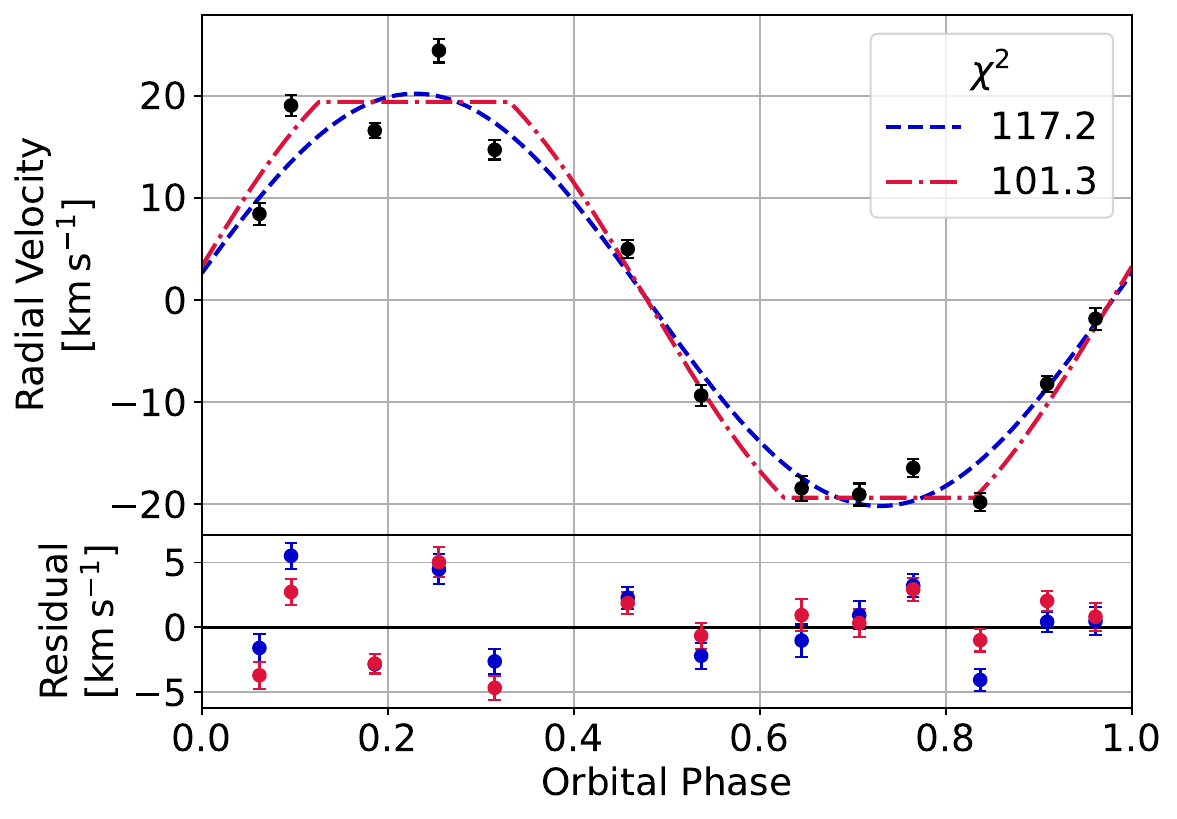}
    \caption{
        Comparison of a simple sine model, an 80\%-clipped sine model, and the donor RV observations from \citet{vdm_2007}.
        The observations were obtained from table 5 of that paper.
        Note that, although this figure is analogous to Figure 9 of that paper, errors are larger in their figure as they were scaled by the reduced $\chi^2$.
        {
            Alt text: This figure consists of two vertical aligned panels.
            The top panel show two sine models and the donor RV observations.
            The bottom panel show residuals for each of the two models.
        }
    }
    \label{fig:vdm_2007}
\end{figure}

Similarly to RV, the weighted average of the donor surface temperature can also be calculated.
This value may not match the temperature determined by spectroscopy or SED fitting, but should serve as a certain representative value for the stellar surface temperature.
The spectral type of SMC\,X-1 has been reported as O9.7 \citep{evans_2004} and B0.2 \citep{lamb_2016} through the spectroscopic observations.
Note that the reported spectral type differs depending on the literature since its optical spectrum is variable.
According to the temperature table of main sequence stars\footnote{\footnoteurl{https:/www.pas.rochester.edu/~emamajek/EEM_dwarf_UBVIJHK_colors_Teff.txt}} \citep{pecaut_2013}, the effective temperature of SMC\,X-1 is estimated as about $30500\text{--}31500\>\Kelvin$.
\zcref[S]{fig:effective_T} shows the weighted-averaged donor surface temperature.
For both $\logq=-1.17, -1.09$ cases, the temperature agrees with that estimated from the spectral type, at the orbital phase where X-ray irradiation is effective.
Furthermore, this figure also shows that the effective temperature of the donor can vary by $1000\text{--}2000\>\Kelvin$ depending on the orbital phase.
Therefore, for determination of the spectral type, the spectroscopic observations should be limited to during the X-ray eclipse.

\begin{figure}
    \includegraphics[width=\linewidth]{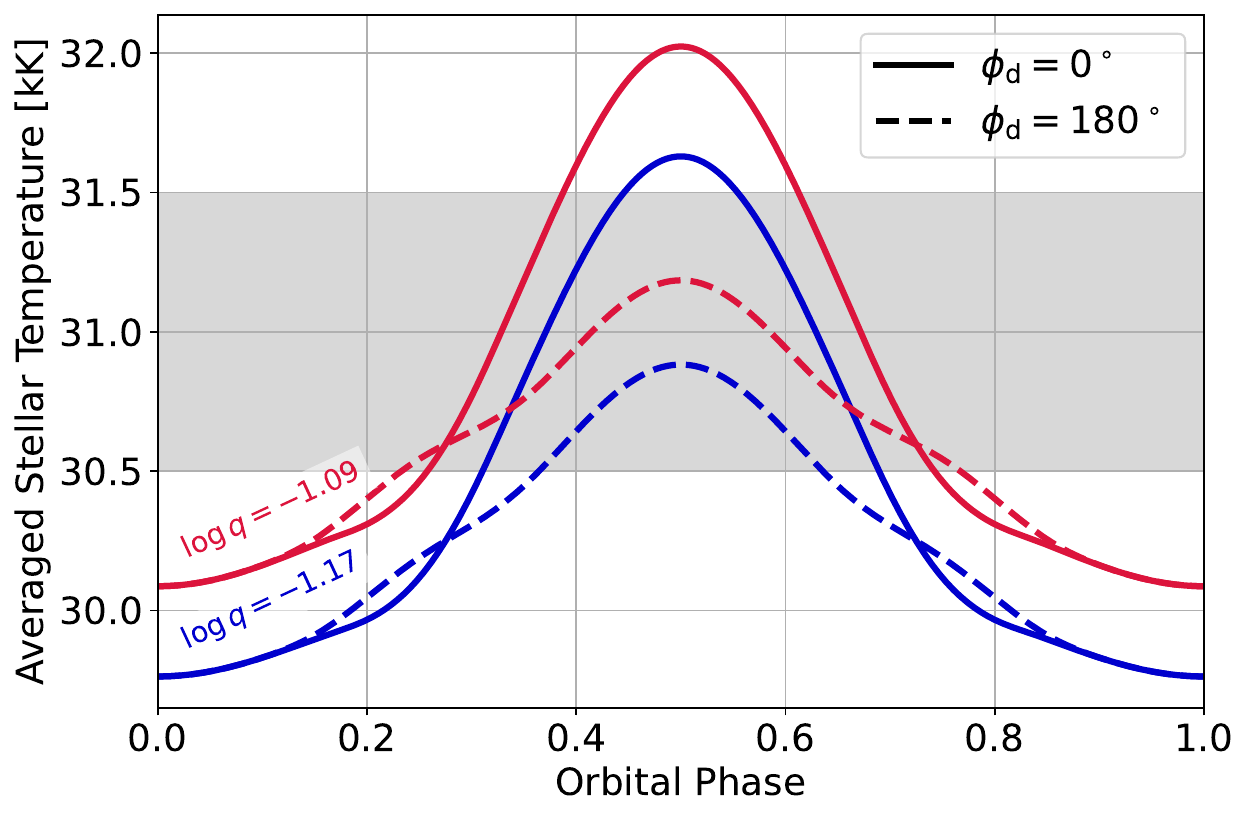}
    \caption{
        Weighted-average temperatures of the donor surface with the parameter-sets of $\logq=-1.17, -1.09$.
        The gray shaded area is a temperature range estimated from the spectral type of SMC\,X-1 ($30500\text{--}31500\>\Kelvin$).
        {
            Alt text: The vertical and horizontal axes show averaged stellar temperature and orbital phase, respectively.
        }
    }
    \label{fig:effective_T}
\end{figure}

\section{Summary}
\label{sec:summary}
We analyzed the ellipsoidal modulation of SMC\,X-1 using \satellite{TESS} and MAXI light curves, for the estimation of its pulsar mass.
The optical light curve of SMC\,X-1 showed two types of notable systematic variations in conjunction with X-ray super-orbital modulation: (1) the minimum at the inferior conjunction is shallow in the off state and vice versa, and (2) the left peak is larger when X-rays are rising, and the right peak is larger when decaying.
While demonstrating these significant optical and X-ray correlations, the absence of optical variations at the superior conjunction (i.e., the pulsar eclipse) provides observational evidence that the origin of these optical variations lies in regions that are occulted during the eclipse (e.g., the accretion disk, the X-ray irradiated surface on the donor).
Based on these observational results, we proposed a modified ellipsoidal modulation model in which the precessing accretion disk changes both the geometry of X-ray irradiation on the donor and that of OUV irradiation on the disk.
Our model succeeded in roughly reproducing the observed optical and X-ray light curves, providing ``optical'' observational evidence for the picture that the precessing disk is the origin of the super-orbital modulation.
Furthermore, we revealed that the donor RV value used in previous mass estimation may be underestimated by about 20\% due to the effect of intense X-ray irradiation on the donor, which shifts the optical emission center away from the gravitational center.
By correcting for this effect, we obtained pulsar mass estimation of about $1.35\>\MO$.

%%%%%%%%%%%%%%%%%%%%%%%%%%%%%%%%%%%%%%%

\begin{ack}
% TESS
This work includes data collected by the \satellite{TESS} mission, and funding for the \satellite{TESS} mission is provided by NASA's Science Mission Directorate.
% MAXI
This research has made use of the MAXI data provided by RIKEN, JAXA, and the MAXI team.
% Gaia
This work has made use of data from the European Space Agency (ESA) mission Gaia\footnote{\footnoteurl{https://www.cosmos.esa.int/gaia}}, processed by the Gaia Data Processing and Analysis Consortium (DPAC\footnote{\footnoteurl{https://www.cosmos.esa.int/web/gaia/dpac/consortium}}). Funding for the DPAC has been provided by national institutions, in particular the institutions participating in the Gaia Multilateral Agreement.
% Python
This work utilized several Python libraries for analysis and visualization: NumPy \citep{haris_2020}, SciPy \citep{virtanen_2020}, Cython \citep{behnel_2011}, CuPy (Okuta et.al. 2017), emcee \citep{foreman_2013}, pandas \citep{mckinney_2010}, and Matplotlib \citep{hunter_2007}.
% Astropy
This work made use of Astropy:\footnote{\footnoteurl{http://www.astropy.org}} a community-developed core Python package and an ecosystem of tools and resources for astronomy \citep{astropy_2013, astropy_2018, astropy_2022}.
\end{ack}

\section*{Funding}
This research was supported by JSPS KAKENHI Grant Number 24H00027.

\section*{Data availability} 
The \satellite{TESS} light curves used in this article were obtained by applying the photometry pipeline of \satellite{TESS} transient project\footnote{\footnoteurl{https://tess.mit.edu/public/tesstransients/}} to full-frame images published in Mikulski Archive for Space Telescopes (MAST) portal\footnote{\footnoteurl{https://archive.stsci.edu/missions-and-data/tess}}.
MAXI GSC $2\text{--}20\,\keV$ light curves we used can be generated using the MAXI on-demand web interface\footnote{\footnoteurl{https://maxi.riken.jp/mxondem}}.

\appendix %%%%%%%%%%%%%%%%%%%%%%%%%%%%%%%%%%%%%%%%%%%%%%%%%%%%%%%%
\section{Procedures of the parabola fitting}\label{sec:parabola_fit}
\begin{figure}
    \includegraphics[width=\linewidth]{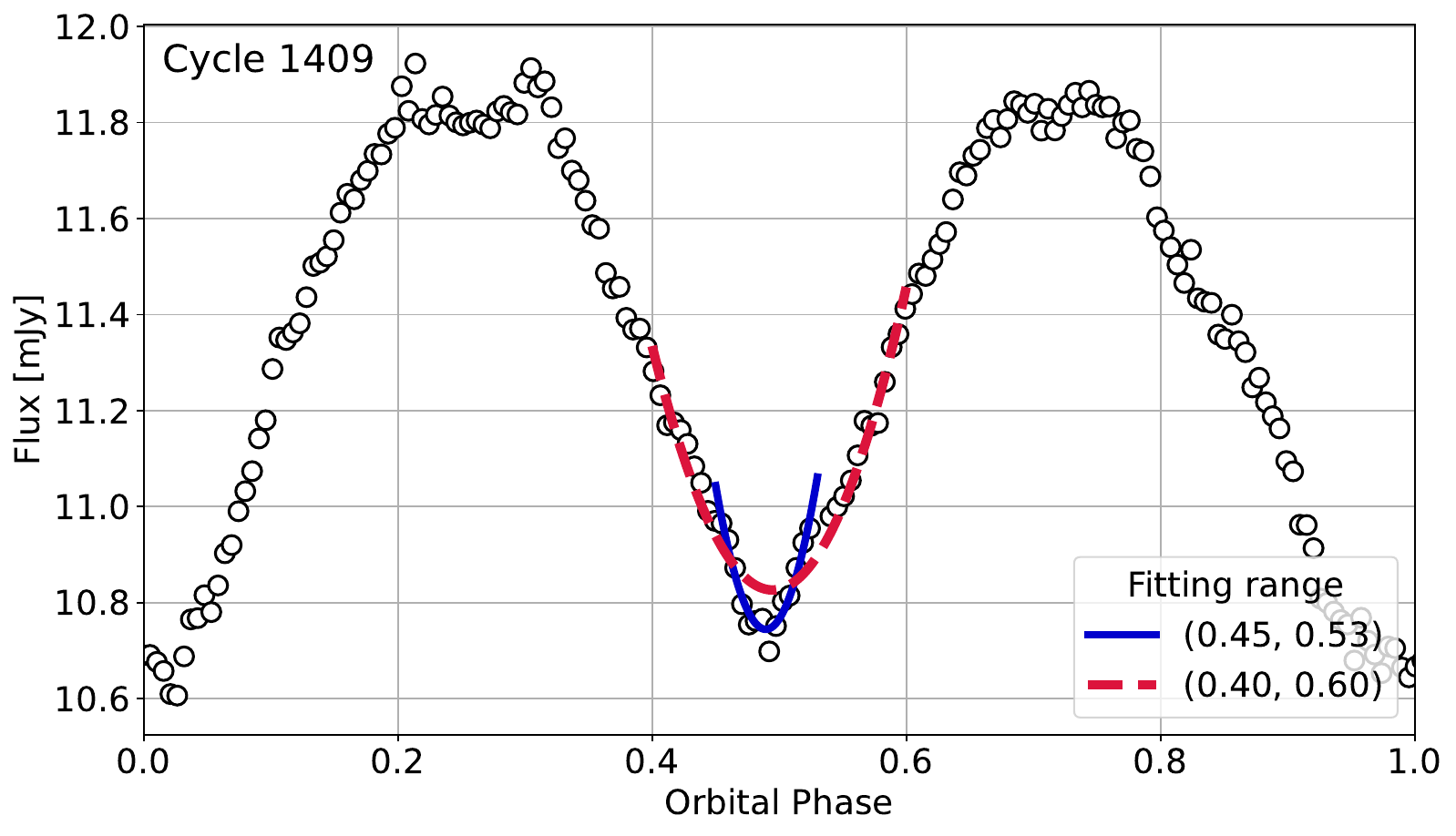}
    \caption{
        An example of differences in parabola-fitting depending on the fitting range.
        In this orbital cycle (1409), the flux decrease around the inferior conjunction occurs in two stages due to the donor occultation by the disk , and if the fitting range is too wide, the minimum flux will be overestimated.
        {
            Alt text: The optical orbital light curve for cycle 1409 and the parabolic fit models of the minimum phase around inferior conjunction are shown for two different fit ranges: $(0.4, 0.6)$ and $(0.45, 0.53)$.
        }
    }
    \label{fig:fit-range}
\end{figure}

Here, we describe the procedure for obtaining the optical fluxes at the peak and minimum phases of the orbital light curve by fitting a parabola.
Since we use a simple and phenomenological model, the choice of the fitting range is important.
In particular, the minimum around the inferior conjunction exhibits a complex light-curve shape due to multiple effects, including changes in the apparent size caused by tidal distortion of the donor, gravitational darkening, X-ray irradiation, and occultation by the disk (\zcref{fig:fit-range}).
Therefore, careful selection of the fitting range is essential.
However, determining the fitting range individually for each orbital cycle is cumbersome, lacks reproducibility, and introduces arbitrariness.
Hence, a unified method for defining the fitting range applicable to all orbital light curves is required.
To address this issue, we adopt an iterative approach: we first perform a preliminary fit, then update the fitting range based on the result, and finally repeat the fitting using the refined range.
The parabolic model is given by:
\begin{equation}
    f=k(\phi-\phi_0)^2+f_0,
\end{equation}
where $f$ is optical flux, $\phi$ is an orbital phase, $\phi_0$ is an orbital phase at the peak or minimum, $k$ is a coefficient of the second-order term, and $f_0$ is the peak or minimum flux.
Initially, fits are performed with an initial fitting range.
Then, the fit is repeated using $(\phi_0-\sqrt{\fw/|k|}, \phi_0+\sqrt{\fw/|k|})$ as the new fitting range, where $\fw$ is a control parameter.
This range update and fitting procedure is performed once more, and the parameters obtained from the final fit are adopted.
We used the same values as initial fitting range and $\fw$ for all orbital cycles.
\zcref[S]{tab:fit-range} lists the values of initial fitting range and $\fw$ for obtaining five peak and minimum fluxes.
Since the phase at the peaks and minima varies depending on the super-orbital phase, the initial fitting range was set to be fairly wide.
The value of $\fw$ was tuned to maximize the fitting range while ensuring that the parabola closely matched the curvature near the peak and minimum.
We visually confirmed that neither the final fitting ranges nor the resulting fits exhibit any significant anomalies for all orbital cycles.

\begin{table}
    \tbl{
        Initial fitting ranges for the parabola fitting
    }{
        \begin{tabular}{cccc}
            \hline
             & \multicolumn{2}{c}{Fitting range} & $w$\\
             & Start & End & [mJy]\\
            \hline
            $\fscf$ & -0.1 & 0.1 & 0.56\\
            $\fleft$ & 0.1 & 0.45 & 0.44\\
            $\fic$ & 0.35 & 0.65 & 0.25\\
            $\fright$ & 0.55 & 0.9 & 0.44\\
            $\fscl$ & 0.9 & 1.1 & 0.56\\
            \hline
        \end{tabular}
    }
    \label{tab:fit-range}
    \begin{tabnote}
    \end{tabnote}
\end{table}

\section{Test of the uniform irradiation approximation}\label{sec:test-singleT}
This section evaluates the impact of the approximation of uniform OUV irradiation onto the disk adopted in our flux calculations (\zcref{sec:model_implement}).
\zcref[S]{fig:tdisk_comparison,fig:fdisk_comparison} compare disk temperature distributions and optical light curves with and without the approximation.
Without the approximation, the disk temperatures show clear non-uniformity, with temperature differences of $1000\text{--}3000\>\Kelvin$ between regions close to and far from the donor depending on the orbital phase.
The difference in disk optical flux by the approximation is within $\approx1\%$ during orbital phases where the disk is not occulted, while the difference increases to $\approx5\%$ during partially occulted phases, with the approximating flux being systematically larger, particularly pronounced when $\phidisk=180^\circ$.
This systematic overestimation arises because, during partially occulted phases, the disk regions closer to the donor are selectively occulted, and these regions are brighter because of abundant OUV scattering and higher temperatures.
Nevertheless, the difference in flux due to the approximation is at most $\sim1\%$ of the total flux variation range, and we concluded that its impact on our results and discussion is negligible.

\begin{figure}
    \includegraphics[width=\linewidth]{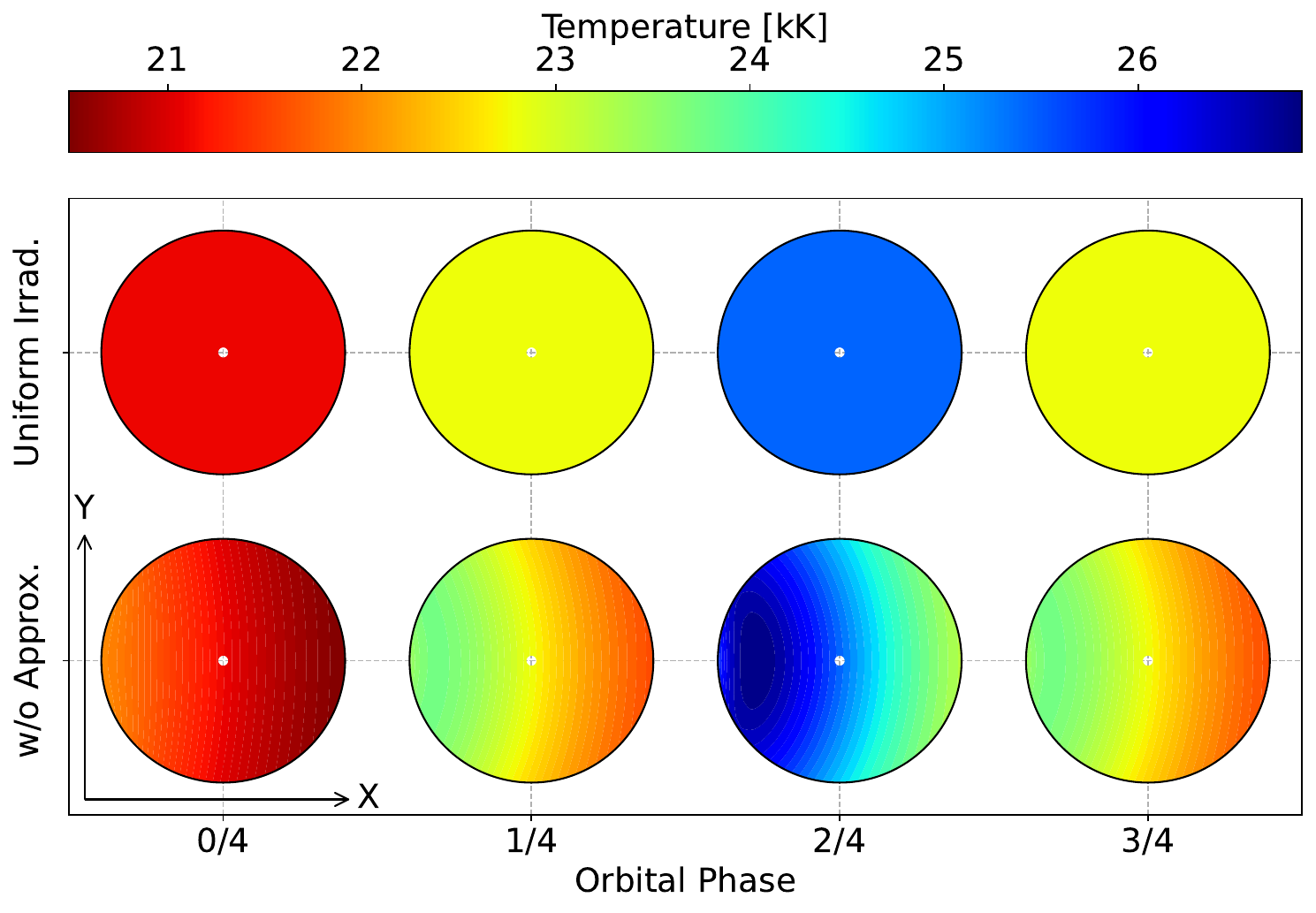}
    \caption{
        Comparison of disk temperature distributions with and without uniform irradiation approximation.
        This shows the case where $\phidisk=0$; for other super-orbital phases, the correspondence between the orbital phase and the disk irradiation geometry is shifted.
        The right side of the figure corresponds to the positive direction of the $X$-axis in the binary cartesian coordinate system (cf. \zcref{sec:modified_ellipmod_model}), and the donor is located to the left of each disk.
        The parameter set used here is for the case of $\logq=-1.17$.
        In the uniform irradiation approximation, the OUV flux from the donor at the pulsar position is applied to the entire disk, so the temperature around the disk center matches that without the approximation.
        {
            Alt text: The disk temperature distribution is shown as filled contours for eight different conditions: four distinctive orbital phases and the presence or absence of approximation.
        }
    }
    \label{fig:tdisk_comparison}
\end{figure}

\begin{figure}
    \includegraphics[width=\linewidth]{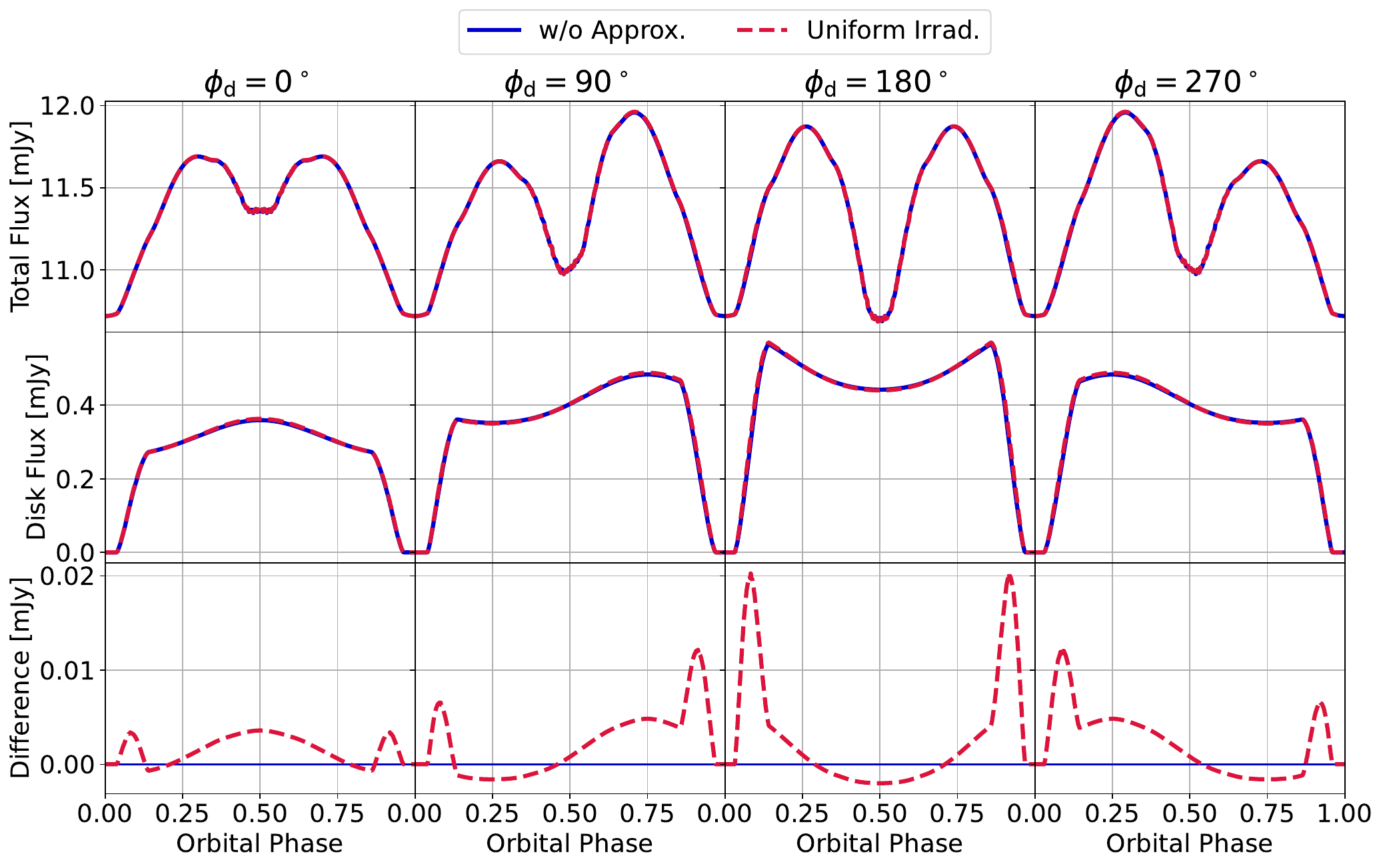}
    \caption{
        Comparison of optical light curves with and without uniform irradiation approximation.
        The bottom panels show the difference between the approximated and the non-approximated disk fluxes.
        The parameter set used here is for the case of $\logq=-1.17$.
        {
            Alt text: This consists of 12 panels arranged in 3 rows and 4 columns.
            The top two rows show light curves for the total flux of the star and disk, and for the disk flux alone, respectively.
            The light curves for the four distinctive super-orbital phases are shown in each of the four columns.
        }
    }
    \label{fig:fdisk_comparison}
\end{figure}

% Any journal's BST file (e.g., apj.bst) can be used as PASJ's BST is unavailable.    
% \bibliographystyle{****}
% \bibliography{****}
\bibliographystyle{references/aasjournal}
\bibliography{
    references/bibfiles/before_2000,
    references/bibfiles/2000-2010,
    references/bibfiles/2010-2020,
    references/bibfiles/after_2020
}

\end{document}